\documentclass[reprint,aps,prb,superscriptaddress]{revtex4-1}
\usepackage{amsmath}
\usepackage{amssymb}
\usepackage{natbib}
\usepackage{graphicx}
\usepackage{xcolor}

\setcitestyle{numbers,square}

\begin{document}

\title{Light scattering by magnons in whispering gallery mode cavities} 
\author{Sanchar Sharma} 
\affiliation{Kavli Institute of NanoScience, Delft University of Technology, 2628 CJ Delft, The Netherlands} 
\author{Yaroslav M. Blanter} 
\affiliation{Kavli Institute of NanoScience, Delft University of Technology, 2628 CJ Delft, The Netherlands} 
\author{Gerrit E. W. Bauer} 
\affiliation{Kavli Institute of NanoScience, Delft University of Technology, 2628 CJ Delft, The Netherlands} 
\affiliation{Institute for Materials Research and WPI-AIMR, Tohoku University, Sendai 980-8577, Japan} 

\begin{abstract}
Brillouin light scattering is an established technique to study magnons, the elementary excitations of a magnet. Its efficiency can be enhanced by cavities that concentrate the light intensity. Here, we theoretically study inelastic scattering of photons by a magnetic sphere that supports optical whispering gallery modes in a plane normal to the magnetization. Magnons with low angular momenta scatter the light in the forward direction with a pronounced asymmetry in the Stokes and the anti-Stokes scattering strength, consistent with earlier studies. Magnons with large angular momenta constitute Damon Eschbach modes are shown to inelastically reflect light. The reflection spectrum contains either a Stokes or anti-Stokes peak, depending on the direction of the magnetization, a selection rule that can be explained by the chirality of the Damon Eshbach magnons. The controllable energy transfer can be used to manage the thermodynamics of the magnet by light.
\end{abstract}

\maketitle 

%%%%%% Intro 

\section{Introduction} 

Magnetic insulators are promising for future spintronics applications such as long-range information transfer \cite{Kajiwara10,Cornelissen15} and low-power logic \cite{Chumak15}. An important representative of this class of materials is yttrium iron garnet (YIG), a ferrimagnetic insulator with very low magnetic damping \cite{Cherepanov_YIG,Serga10,WuHoff}. The elementary magnetic excitations (\emph{magnons}) in YIG have long coherence times \cite{ZhangMemory15}, enabling the study of ``quantum magnonics'' \cite{TabuchiQubit15,Tabuchi16}. From this perspective, it is interesting to develop methods for classical, and eventually quantum, magnon manipulation. 

Magnons are known to couple to a wide range of carriers such as electrons \cite{Silsbee79,Brataas12}, phonons \cite{Kittel58,ZhangPhonon15}, microwaves \cite{SoykalPRL10,Huebl13}, and light \cite{EllLou}. Recently, experimental progress has been reported in the coupling of YIG spheres to optical and microwave photons. Microwaves can coherently exchange information with magnons, in both quantum (Rabi oscillatory) \cite{SoykalPRL10,TabuchiHybrid14,Hisatomi16} and classical \cite{Zhang14,Bai15,Nikita15,Bourhill16} regimes. Magnons can also coherently couple to photons at optical frequencies \cite{Tianyu16,Hisatomi16,Silvia16} via a two-photon scattering mediated by the magnetization. For efficient magnon manipulations the magnon-photon coupling must be enhanced, which is possible by confining photons and magnons in a cavity. By analogy with cavity optomechanics \cite{Aspelmeyer14}, in which cavity photons interact with mechanical degrees of freedom, this field has been dubbed \emph{cavity optomagnonics}. This optomagnonic interaction can be used to selectively pump magnons \cite{ZhangWGM16,Osada16}, where the magnet serves as an optical cavity. First theoretical papers in cavity optomagnonics recently emerged, which emphasize the possibility and importance of strong interaction between magnetism and light. Coherent effects such as electromagnetically induced transparency and Purcell effect in planar cavities \cite{Tianyu16}, as well as optically generated magnetization dynamics in spheres \cite{Silvia16} have been proposed. 

Optomagnonic interactions cause elastic or inelastic light scattering and were studied in bulk materials for a long time, both experimentally \cite{Pisarev70,Borovik82,Vedernikov87} and theoretically \cite{BassKag,EllLou,Hu_Morgen,Wettling75,BorovikSinha}. The elastic scattering caused by magnetically induced birefringence causes the Faraday (Fa) and Cotton-Mouton (CM) effects \cite{WettlingData76,Kirilyuk10}, collectively known as magneto-optical (MO) effects. They have also been experimentally studied in cavities \cite{Haigh15}. 

The inelastic magnetic scattering is observable in Brillouin light scattering (BLS) \cite{WettlingData76}, in which photons exchange energy with the magnetization by creating or annihilating magnons. BLS in which photons lose or gain energy is referred to as Stokes (S) or anti-Stokes (aS) scattering, respectively. In conventional BLS spectroscopy, the change in the momentum and energy of the photons is used to measure the dispersion and spatial profile of magnons \cite{Sebastian15Rev}. 

When BLS occurs in a cavity, the incident and scattered photons are confined by its boundaries, which may increase the scattering efficiency. Cavities with ellipsoidal geometries such as spheres or disks support whispering gallery modes (WGMs) \cite{Richtmyer} that can exhibit high Q-factors \cite{OraevskyWGM,Yang15}. WGMs have found various applications in optical engineering \cite{Foreman15,Yang15} and the study of light-matter interaction \cite{OraevskyWGM,Yang15}. WGMs can be pictured in terms of consecutive total internal reflection on a curved surface with closed orbits \cite{Yang15}. 

BLS in a WGM cavity made from a magnetic material displays a pronounced asymmetry in the Stokes and anti-Stokes light scattering intensities \cite{ZhangWGM16,Osada16,HaighWGM}. Such an S-aS asymmetry has been observed in other magnetic systems too, e.g. due to an interference of photons affected by different microscopic scattering mechanisms \cite{Wettling75,BorovikSinha,Lockwood12}. An another source for S-aS imbalance is an ellipticity of the spin waves that is caused by magnetic anisotropies \cite{Camley82}. This asymmetry is observed in thick films too \cite{Grunberg77,Grimsditch79} due to the asymmetric localization of Damon-Eshbach (DE) modes on one of the surfaces \cite{DamEshSurface,DamEshSlab}. The S-aS asymmetry in WGM cavities was attributed to quite different phenomena, viz. the partial elliptical polarization of WGMs \cite{ZhangWGM16,Osada16} or the interplay of birefringence and conservation laws \cite{HaighWGM}. 

\begin{figure}[ptb]
\begin{equation*}
\includegraphics[width=.47\textwidth,keepaspectratio]{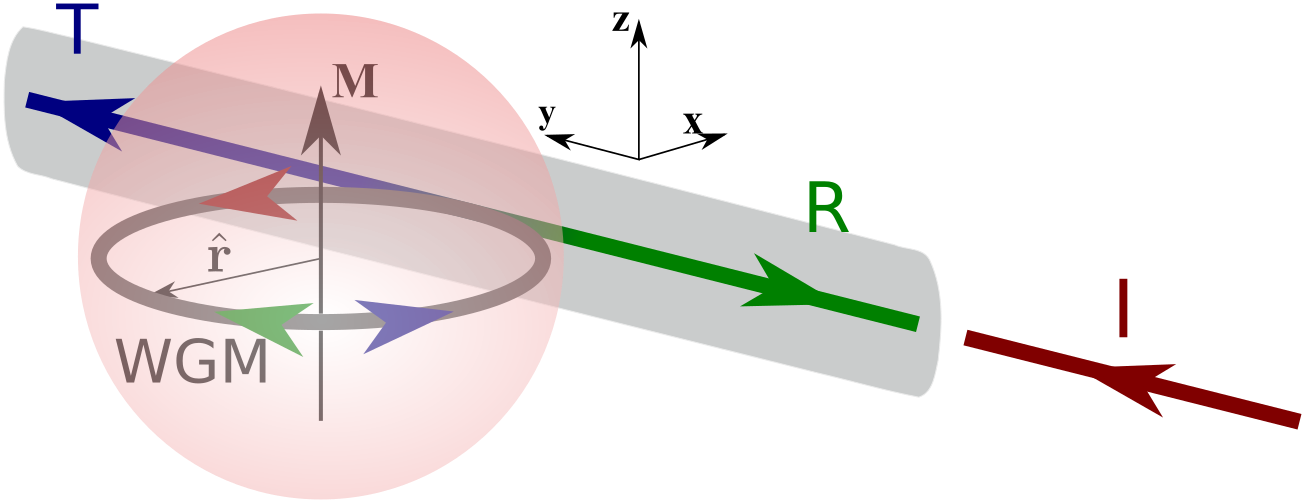}
\end{equation*}
\caption{A magnetic sphere is coupled to an evanescent coupler which can
excite optical modes inside the sphere. Incident photons (I) in the coupler
scatter inelastically by the magnetic order in the sphere, which can be
observed in the reflected (R) and transmitted (T) light that leads back into
the coupler. The corresponding counter-propagating WGMs are shown by green (R)
and blue (T) colored arrows inside the sphere.}
\label{Setup}
\end{figure}

Here we theoretically study light scattering by magnons in magnetic spheres in which the WGMs are excited by evanescent coupling to a light source, such as an illuminated waveguide, a tapered fibre or a prism. We generalize previous works by including all the magnons which contribute significantly to BLS. In particular, we differentiate between the transmission and reflection in the coupler attributed to different magnons. We consider magnetic spheres with sub-mm radii as shown in Fig. \ref{Setup}. The magnetization is assumed to be saturated by an external magnetic field. 

We consider the power spectrum of inelastically transmitted and reflected spectra for a given input light source, for both Stokes and anti-Stokes photons, emphasizing the S-aS asymmetry. We present analytic results for specific magnons and provide estimates for the other magnons. We find a pronounced S-aS asymmetry in the transmission, as observed in recent experiments for the Kittel mode \cite{Osada16,ZhangWGM16,HaighWGM}. Our theory agrees with and generalizes the analysis of \cite{HaighWGM}. Very recently, the transmission due to other (``Walker'') magnons have been observed as well \cite{UsamiUnpub}. We predict that photons are inelastically reflected by DE magnons with complete suppression of either Stokes or anti-Stokes lines. The latter results can be interesting for thermodynamic applications. 

This manuscript is organized as follows. We start with introducing the observables and qualitative considerations in Sec. \ref{Sec:Qual}. We calculate the transmitted and the reflected power for a general cavity coupled to an evanescent coupler (a proximity optical fiber) in Sec. \ref{Sec:Output}. We introduce the details of the model in Sec. \ref{Sec:Model}, recapitulating basic concepts of WGMs and magnons from the literature. We calculate the scattering amplitude of WGMs in Sec. \ref{Sec:Coup}. We discuss the physical consequences of the theory by considering an example of a YIG sphere with a particular input in Sec. \ref{Sec:TxRx}. We generalize the treatment of Sec. \ref{Sec:TxRx} to other input modes in Sec. \ref{Sec:Other}. We summarize results and give an outlook in Sec. \ref{Sec:Conc}. 

%%%%%%%Qualitative Discussion

\section{Initial considerations} \label{Sec:Qual}

We first discuss our setup shown in Fig. \ref{Setup} and a few qualitative aspects to set the stage. An optical waveguide guides the incoming and outgoing (near-infrared) light radiation along the $\pm y$-axis. We assume that the waveguide is a thin optical fiber that supports only one transverse optical mode (single mode fiber) with two polarization components corresponding to $\mathbf{E}\parallel\hat{\mathbf{z}}$ or $\mathbf{E} \parallel\hat{\mathbf{x}}$, which we label as transverse electric ($\varsigma = \mathrm{TE}$) and transverse magnetic ($\varsigma = \mathrm{TM}$), respectively. The power spectrum of each polarization component is denoted by $P_{\mathrm{in}}^{\varsigma}(\omega)$. The waveguide is optically coupled to the magnetic sphere due to the overlap of the transverse evanescent light amplitudes. We focus on the optical coupling to a single-mode fiber, but application to other geometries such as an attached prism or multi-mode wave guide is straightforward. The output power spectrum addressed here has three components: (1) the transmission without coupling with the magnons, $P_{\mathrm{el}}^{\varsigma}(\omega)$; (2) the light scattered by magnons in the forward direction that can be observed in transmission, $P_{\mathrm{T}}^{\varsigma}(\omega)$; (3) the light reflected by magnons, $P_{\mathrm{R}}^{\varsigma}(\omega)$. 

The incident photons predominantly excite the WGMs of the sphere, which are optical modes confined to the equatorial surface and thus have the largest overlap with the optical waveguide modes \cite{Chiasera10}. The WGMs of large spheres are to a good approximation linearly TM or TE polarized plane EM waves that propagate adiabatically along the equator. The coupling to the optical waveguide is assumed to conserve the polarization, which is a good approximation for clean contacts. 

The excited WGMs may be scattered by magnons into a multitude of other optical modes mediated by the optomagnonic interaction, to be discussed in detail below. We only consider scattering among the WGMs, because only those couple to the optical waveguide and lead to observable effects. We take the magnetization to be along $ + z$-axis (and later also consider the case of $ - z$ ). In this configuration, elastic scattering effects mediated by the magnetization are symmetry-forbidden \cite{BorovikSinha,Wettling75}, which simplifies the analysis. We treat the optomagnonic interaction by perturbation theory, which is valid if the magnons are not significantly affected by light. The frequency of optical photons is several orders of magnitude larger than that of magnons, so the incident and the scattered light has almost the same frequency. This implies that the scattering between the WGMs to a good approximation preserves the modulus of the (azimuthal) momentum, while a reversal of the direction (reflection or backscattering) is allowed \cite{Sebastian15Rev}. The forward and backward scattered light are indicated by the blue and green arrows in Fig. \ref{Setup}, respectively. The scattered WGMs leak back into the optical waveguide, propagating towards $ + y$ or $ - y$ depending on the circulation sense of WGMs [as shown by the color-matched arrows in Fig. \ref{Setup}]. 

Since forward scattering nearly conserves photon momentum, the involved magnons must have small angular momentum, denoted here as small-L (SL) magnons. The most important SL magnon is the uniformly precessing magnetization (macrospin or Kittel) mode with zero orbital angular momentum. In contrast, the reflection of photons is caused by magnons whose angular momentum is approximately twice of that of the incident WGM. High angular momentum transfer can be provided by the Damon-Eshbach (DE) surface modes localized to the equator of the sphere \cite{DamEshSurface}. These magnons are chiral, viz. they always circulate in an anti-clockwise fashion w.r.t the magnetization (parallel to the blue arrow in Fig. \ref{Setup}). We denote the set of DE magnons as large-L (LL) magnons. 

Before going into the details of the scattering mechanism inside the sphere, we discuss the photon transport in the single-mode optical fiber evanescently coupled to the sphere. The equations can be easily carried over to discuss coupling via a prism attached to the sphere \cite{HaighWGM}. 

%%%%Scattering Matrices

\section{Output spectrum} \label{Sec:Output} 

Here we derive the power spectrum of transmitted and reflected photons for a given source by the input-output (IO) formalism \cite{Gardiner_Orig,Clerk10} . Consider an incident light beam with polarization dependent amplitude $\hat{A}_{\mathrm{in}}^{\varsigma}(t)$, where $\varsigma\in\{\mathrm{TE}, \mathrm{TM}\}$. $\hat{A}_{\mathrm{in}}^{\varsigma}$ is the annihilation operator of the incoming traveling photons that satisfy the commutation rules $\left[ \hat{A}_{\mathrm{in}}^{\varsigma}(t^{\prime}),\hat{A}_{\mathrm{in}}^{\varsigma^{\prime}\dagger}(t)\right] = \delta_{\varsigma,\varsigma^{\prime}}\delta(t - t^{\prime})$. The overlap of the fiber and WGM modes generates photons in the sphere. The latter are expressed by the annihilation operators $\{\hat{a}_{\mathbf{p}}\}$ for a mode $\mathbf{p}$ (discussed for WGMs in Sec \ref{Sec:WGM}). The $\hat{H}$ Hamiltonian for the sphere as derived in Section \ref{Sec:Model} leads to the equation of motion 
\begin{equation}
	 \frac{d\hat{a}_{\mathbf{p}}}{dt} = \frac{i}{\hbar}\left[ \hat{H},\hat{a}_{\mathbf{p}}\right] - \frac{\bar{\kappa}_{\boldsymbol{\mathbf{p}}}}{2}\hat{a}_{\mathbf{p}} - \sum_{\varsigma}\left( \frac{\kappa_{\mathbf{p}}^{\varsigma}}{2} \hat{a}_{\mathbf{p}} + \sqrt{\kappa_{\mathbf{p}}^{\varsigma}}\hat{A}_{\mathrm{in}}^{\varsigma}\right) ,\label{apEq} 
\end{equation} 
where $\bar{\kappa}_{\boldsymbol{\mathbf{p}}}$ is the intrinsic dissipation rate of mode $\mathbf{p}$ in the sphere. $\kappa_{\boldsymbol{\mathbf{p}}}^{\varsigma}$ parameterizes the coupling between the fiber and the WGMs \cite{Gardiner_Orig,Clerk10} via the term $\sqrt{2\kappa_{\mathbf{p}}^{\varsigma}}\hat{A}_{\mathrm{in}}^{\varsigma}$ as well as the dissipation by the escape of $\varsigma$-polarized WGM photons into the fiber. 

The IO formalism relates the input and output amplitudes as 
\begin{equation}
	 \hat{A}_{\mathrm{out}}^{\varsigma} = \hat{A}_{\mathrm{in}}^{\varsigma} + \sum_{\mathbf{p}}\sqrt{\kappa_{\boldsymbol{\mathbf{p}}}^{\varsigma}}\hat{a}_{\mathbf{p}},\label{IOBasic} 
\end{equation} 
where $\hat{A}_{\mathrm{out}}^{\varsigma}$ includes both transmitted and reflected photons. As discussed above, we can separate the transmitted and reflected components based on the circulation sense of WGMs which is coded in the mode index $\mathbf{p}$ [see Fig. \ref{Setup}]. $\hat{A}_{\mathrm{out}}^{\varsigma}$ governs the light observables in the fiber after interaction with the sphere. Eqs. (\ref{apEq}) and (\ref{IOBasic}) leads to $\hat{A}_{\mathrm{out}}^{\varsigma}$ in terms of $\hat{A}_{\mathrm{in}}^{\varsigma}$. 

We now relate the amplitudes $\hat{A}_{\mathrm{in}}^{\varsigma}$ and $\hat{A}_{\mathrm{out}}^{\varsigma^{\prime}}$ to the corresponding power spectra $P_{\mathrm{in}}^{\varsigma}(\omega)$ and $P_{\mathrm{out}}^{\varsigma^{\prime}}(\omega)$ respectively. The power spectrum $P$ corresponding to a field operator $\hat{A}$ can be expressed as \cite{Clerk10} 
\begin{equation}
	 \hbar\omega\left\langle \hat{A}^{\dagger}(\omega_1)\hat{A} (\omega )\right\rangle = 2\pi P(\omega)\delta(\omega + \omega_1),\label{Def:Pow} 
\end{equation} 
where the Fourier transform is defined as 
\begin{equation}
	 \hat{f}(\omega)\overset{\triangle}{=}\int dt\ e^{i\omega t}\hat{f}(t) . 
\end{equation} 

By replacing $\hat{A}\rightarrow\hat{A}_{\mathrm{in}}^{\varsigma},\hat{A}_{\mathrm{out}}^{\varsigma}$ and $P\rightarrow P_{\mathrm{in}}^{\varsigma},P_{\mathrm{out}}^{\varsigma}$ respectively, we can define the input and output power spectra. We find $P_{\mathrm{out}}$ in terms of $P_{\mathrm{in}}$ via the amplitude $\hat{A}_{\mathrm{out}}^{\varsigma}$ in terms of $\hat{A}_{\mathrm{in}}^{\varsigma}$. 

\subsection{Output amplitude} \label{Sec:OAmp}
 
We first discuss the frequency dependence of the operators from which the power spectra can be obtained using Eq. (\ref{Def:Pow}). We treat the magnetism in terms of non-interacting spin waves or magnons, which is valid in the limit of small density and/or long wavelength of magnons. The Hamiltonian for the sphere derived in Section \ref{Sec:Model} can be written 
\begin{equation}
	 \hat{H} = \sum_{\mathbf{p}}\hbar\omega_{\mathbf{p}}\hat{a}_{\mathbf{p}}^{\dagger}\hat{a}_{\mathbf{p}} + \sum_{\boldsymbol{\alpha}}\hbar \omega_{\boldsymbol{\alpha}}\hat{c}_{\boldsymbol{\alpha}}^{\dagger}\hat{c}_{\boldsymbol{\alpha}} + \hat{H}_{\mathrm{OM}},\label{SQHam} 
\end{equation} 
where the $\hat{c}_{\boldsymbol{\alpha}}$ are annihilation operators of magnon in the sphere with mode index $\boldsymbol{\alpha}$, $\omega_{\mathbf{p}} $($\omega_{\boldsymbol{\alpha}}$) are the photon (magnon) frequencies, and $\hat{H}_{\mathrm{OM}}$ represents the optomagnonic interaction. 

Since the optomagnonic interaction is weak, we can expand $\hat{H}_{\mathrm{OM}}$ to leading order in the possible scattering processes. The photonic and magnonic modes have typical frequencies $\omega_{\mathbf{p}} \sim100 - 1000$ THz and $\omega_{\boldsymbol{\alpha}}\sim1 - 10$ GHz, respectively. Optical absorption $\sim\hat{a}_{\mathbf{p}}^{\dagger}\hat{c}_{\boldsymbol{\alpha}}$ and two-photon generation $\sim\hat{a}_{\mathbf{p}}^{\dagger}\hat{a}_{\mathbf{q}}^{\dagger}\hat{c}_{\boldsymbol{\alpha}}$ can be safely disregarded since $\omega_{\boldsymbol{\alpha}}\ll\omega_{\mathbf{p}}$. The leading interaction terms are of the light-scattering form 
\begin{equation}
	 \hat{H}_{\mathrm{OM}} = \hbar\sum_{\mathbf{pq}\boldsymbol{\alpha}}\hat{a}_{\mathbf{p}}\hat{a}_{\mathbf{q}}^{\dagger}\left( G_{\mathbf{pq} \boldsymbol{\alpha}}^+\hat{c}_{\boldsymbol{\alpha}} + G_{\mathbf{pq} \boldsymbol{\alpha}}^-\hat{c}_{\boldsymbol{\alpha}}^{\dagger}\right) . \label{Hint} 
\end{equation} 

$G_{\mathbf{pq}\boldsymbol{\alpha}}^+$ parameterizes, e.g., the amplitude for the scattering of an WGM from mode $\mathbf{p}$ into $\mathbf{q}$ by annihilating an $\boldsymbol{\alpha}$-magnon. We derive expressions for these matrix elements in the sections below for spherical samples. 

Inserting Eq. (\ref{Hint}) into Eq. (\ref{apEq}) leads to the coupled operator equation, 
\begin{align}
	 \hat{a}_{\mathbf{q}}\left( \omega\right) &= - \chi_{\mathbf{q}} (\omega)\left\{\sum_{\varsigma}\sqrt{\kappa_{\boldsymbol{\mathbf{q}}}^{\varsigma}}\hat{A}_{\mathrm{in}}^{\varsigma}\left( \omega\right) \right . \nonumber \\
	 & + \left . \sum_{\boldsymbol{\mathbf{p\alpha}}}\left[ \hat{a}_{\mathbf{p}}\ast\left( G_{\boldsymbol{\mathbf{pq\alpha}}}^+\hat{c}_{\boldsymbol{\alpha}} + G_{\boldsymbol{\mathbf{pq\alpha}}}^-\hat{c}_{\boldsymbol{\alpha}}^{\dagger}\right) \right] (\omega)\right\} ,\label{Eq:aqSol} 
\end{align} 
where 
\begin{equation}
	 \chi_{\mathbf{q}}(\omega) = \frac{1}{- i(\omega - \omega_{\mathbf{q}}) + (\bar{\kappa}_{\boldsymbol{\mathbf{q}}} + \kappa_{\boldsymbol{\mathbf{q}}}^{\mathrm{TE}} + \kappa_{\boldsymbol{\mathbf{q}}}^{\mathrm{TM}} )/2},\label{Def:chi} 
\end{equation} 
is the susceptibility of the $\mathbf{q}$-WGM, and $\ast$ denotes convolution in the frequency domain, 
\begin{equation}
	 \left[ \hat{f}\ast\hat{g}\right] (\omega)\overset{\triangle}{=}\int \frac{d\omega^{\prime}}{2\pi}\hat{f}\left( \omega^{\prime}\right) \hat{g} \left( \omega - \omega^{\prime}\right) . 
\end{equation} 
To leading order in $G^{\pm}$, we may linearise the equation $\hat{a}_{\mathbf{p}}\rightarrow - \sum_{\varsigma}\chi_{\mathbf{p}}\sqrt{\kappa_{\boldsymbol{\mathbf{p}}}^{\varsigma}}\hat{A}_{\mathrm{in}}^{\varsigma}$ on the right-hand-side of Eq. (\ref{Eq:aqSol}). Its solution for $\hat{a}_{\mathbf{q}}$ can be inserted into Eq. (\ref{IOBasic}) leading to the output amplitude 
\begin{equation}
	 \hat{A}_{\mathrm{out}}^{\varsigma^{\prime}}(\omega) = \hat{A}_{\mathrm{el}}^{\varsigma^{\prime}}(\omega) + \hat{A}_{\mathrm{T}}^{\varsigma^{\prime}}( \omega) + \hat{A}_{\mathrm{R}}^{\varsigma^{\prime}}(\omega) . \label{Aout:Full} 
\end{equation} 
The contribution, $\hat{A}_{\mathrm{el}}^{\varsigma^{\prime}}$ describes the purely dielectric and elastic response, i.e. the transmission without optomagnonic coupling, $G^{\pm} = 0$ 
\begin{equation}
	 \hat{A}_{\mathrm{el}}^{\varsigma^{\prime}}(\omega) = \hat{A}_{\mathrm{in}}^{\varsigma^{\prime}}(\omega) - \sum_{\mathbf{q},\varsigma}\chi_{\mathbf{q}}(\omega)\sqrt{\kappa_{\boldsymbol{\mathbf{q}}}^{\varsigma^{\prime}} \kappa_{\boldsymbol{\mathbf{q}}}^{\varsigma}}\ \hat{A}_{\mathrm{in}}^{\varsigma}(\omega) . 
\end{equation} 
The photons forward or backward scattered by the magnons are given by $\hat{A}_{\mathrm{T}}$ and $\hat{A}_{\mathrm{R}}$ respectively, where 
\begin{align}
	 \hat{A}_{\mathrm{T}}^{\varsigma^{\prime}}(\omega^{\prime}) &= \sum_{\substack{\boldsymbol{\mathbf{pq}},\varsigma\\
	\boldsymbol{\alpha} \in\text{\textrm{SL}}}}\int\frac{d\omega}{\pi}\sqrt{\kappa_{\boldsymbol{\mathbf{q}}}^{\varsigma^{\prime}}\kappa_{\boldsymbol{\mathbf{p}}}^{\varsigma}}\chi_{\mathbf{q}}(\omega^{\prime})\chi_{\mathbf{p}}(\omega )\hat{A}_{\text{\textrm{in}}}^{\varsigma}(\omega) \nonumber \\ & \left[ G_{\boldsymbol{\mathbf{pq\alpha}}}^+ \hat{c}_{\boldsymbol{\alpha}}\left( \omega^{\prime} - \omega\right) + G_{\boldsymbol{\mathbf{pq\alpha}}}^- \hat{c}_{\boldsymbol{\alpha}}^{\dagger}(\omega^{\prime} - \omega) \right] ,\label{AT:Gen} 
\end{align} 
and a similar equation is given by the replacements $\hat{A}_{\mathrm{T}}\rightarrow\hat{A}_{\mathrm{R}}$ and S$\text{L}\rightarrow\text{LL , where}$ S$\text{L}$ and L$\text{L}$ are the set of small-L and large-L magnons, as explained above. 

We can interpret Eq. (\ref{AT:Gen}) in terms of the following scattering processes 
\begin{equation}
	 \hat{A}_{\mathrm{in}}^{\varsigma}(\omega)\rightarrow\hat{a}_{\mathbf{p}} \xrightarrow{\hat{c}_{\boldsymbol{\alpha}}\in\text{SL}}\hat{a}_{\mathbf{q}}\rightarrow\hat{A}_{\mathrm{T}}^{\varsigma^{\prime}}(\omega^{\prime}) . 
\end{equation} 

The incoming photons with polarization $\varsigma$ at frequency $\omega$ couple to the $\mathbf{p}$-WGMs with rate $\propto\sqrt{\kappa_{\boldsymbol{\mathbf{p}}}^{\varsigma}}\chi_{\mathbf{p}}(\omega)$. Each of the $\mathbf{p}$ -modes is scattered by a small-L $\boldsymbol{\alpha}$ -magnon to a $\mathbf{q}$-WGM with rate $\propto G_{\boldsymbol{\mathbf{pq\alpha}}}^{\pm}$. The scattered $\mathbf{q}$-WGMs are transferred back into the fiber with polarization $\varsigma^{\prime}$ and frequency $\omega^{\prime}$ at rates $\propto\sqrt{\kappa_{\boldsymbol{\mathbf{q}}}^{\varsigma^{\prime}}} \chi_{\mathbf{q}}(\omega^{\prime})$. Summing over all $\boldsymbol{\mathbf{pq\alpha}}$ gives the output as a function of input frequency and polarization. A similar equation involving large-$L$ magnons gives the reflected amplitude. 

\subsection{Output power} \label{Sec:OPow}

Eq. (\ref{Aout:Full}) can be used to derive the output power spectrum $P_{\mathrm{out}}$ in terms of the expectation value in Eq. (\ref{Def:Pow}) involving squared $\hat{A}_{\mathrm{out}}$. We assume that the TE and TM polarized components of input light are uncorrelated, 
\begin{equation}
	 \left\langle \left( \hat{A}_{\mathrm{in}}^{\varsigma^{\prime}}\right)^{\dagger}(\omega_1)\hat{A}_{\mathrm{in}}^{\varsigma}(\omega)\right\rangle = 0,\ \varsigma\neq\varsigma^{\prime} . 
\end{equation} 
This is valid if the input is TE or TM polarized. The auto-correlation function of $\hat{A}_{\mathrm{in}}^{\varsigma}$ defines the input power according to Eq. (\ref{Def:Pow}). Since magnons are only weakly perturbed by the light, so we have $\left\langle \hat{c}_{\boldsymbol{\alpha}}^{\dagger} \hat{A}_{\mathrm{in}}\right\rangle = \left\langle \hat{c}_{\boldsymbol{\alpha}}\hat{A}_{\mathrm{in}}\right\rangle = 0$. Therefore the elastically scattered light $\hat{A}_{\mathrm{el}}$ does not interfere with $\hat{A}_{\mathrm{T}}$ and $\hat{A}_{\mathrm{R}}$, i.e. 
\begin{equation}
	 \left\langle \hat{A}_{\mathrm{el}}\hat{A}_{\mathrm{T}}\right\rangle = \left\langle \hat{A}_{\mathrm{el}}\hat{A}_{\mathrm{R}}\right\rangle = 0 . 
\end{equation} 

In contrast to the photons, the magnons are at ambient temperatures thermally occupied even without external stimulation. Thermal equilibrium of magnons can be modelled by the interactions with a memory-less (Markovian) bath at temperature $T$ \cite{Clerk10} that consists of a quasi-continuum of bosonic oscillators $\{\hat{B}_{\boldsymbol{\Omega}}\},$ where $\Omega$ is the frequency of an oscillator in mode $\boldsymbol{\Omega}$ and annihilation operator $\hat{B}_{\boldsymbol{\Omega}}$ with $\left[ \hat{B}_{\boldsymbol{\Omega}},\hat{B}_{\boldsymbol{\Omega}^{\prime}}^{\dagger}\right] = \delta_{\boldsymbol{\Omega\Omega}^{\prime}}$. The equation of motion of an $\boldsymbol{\alpha}$-magnon can then be written as 
\begin{equation}
	 \frac{d\hat{c}_{\boldsymbol{\alpha}}(t)}{dt} = - i\omega_{\boldsymbol{\alpha}} \hat{c}_{\boldsymbol{\alpha}}(t) - \frac{\bar{\kappa}_{\boldsymbol{\mathbf{\alpha}}}}{2}\hat{c}_{\boldsymbol{\alpha}}(t) - \sqrt{\bar{\kappa}_{\boldsymbol{\mathbf{\alpha}}}}\hat{b}_{\boldsymbol{\alpha}}(t), 
\end{equation} 
where $\bar{\kappa}_{\boldsymbol{\mathbf{\alpha}}}$ is the intrinsic linewidth that in the model below reads $\bar{\kappa}_{\boldsymbol{\alpha}} = \alpha_G\omega_{\boldsymbol{\alpha}}$ in terms of the Gilbert damping $\alpha_G $. $\hat{b}_{\boldsymbol{\alpha}}$ represents a fluctuating noise source acting on the $\boldsymbol{\alpha}$-magnon and generated by the bath. It is given approximately by \cite{Clerk10}, 
\begin{equation}
	 \hat{b}_{\boldsymbol{\alpha}}(t) \approx \frac{1}{\sqrt{2\pi\rho(\omega_{\boldsymbol{\alpha}})}}\sum_{\overset{\boldsymbol{\Omega}}{\left\vert \Omega - \omega_{\boldsymbol{\alpha}}\right\vert <\bar{\kappa}_p}}\hat{B}_{\boldsymbol{\Omega}}(t_0)e^{- i\Omega(t - t_0)}, 
\end{equation} 
where $t_0\rightarrow - \infty$ is some initial time, $\rho(\omega_{\boldsymbol{\alpha}})$ is the density of states of the bath at frequency $\omega_{\boldsymbol{\alpha}}$ [see \cite{RivasHuelga} for a proper mathematical treatment]. The bath operators with $\left\langle \hat{b}_{\boldsymbol{\alpha}}(t)\right\rangle = 0$ are assumed to obey the commutation rules 
\begin{equation}
	 \left[ \hat{b}_{\boldsymbol{\alpha}}(t^{\prime}),\hat{b}_{\boldsymbol{\beta}}^{\dagger}(t)\right] = \delta_{\boldsymbol{\alpha}\boldsymbol{\beta}} \delta(t - t^{\prime}) . \label{Eq:MagBathComm} 
\end{equation} 

At equilibrium 
\begin{equation}
	 \left\langle \hat{b}_{\boldsymbol{\alpha}}^{\dagger}(t^{\prime})\hat{b}_{\boldsymbol{\beta}}(t)\right\rangle = \delta_{\boldsymbol{\alpha}\boldsymbol{\beta}}n_{\boldsymbol{\alpha}}\delta(t - t^{\prime}), 
\end{equation} 
where $n_{\boldsymbol{\alpha}} = \left( \exp\frac{\hbar\omega_{\boldsymbol{\alpha}}}{k_BT} - 1\right)^{- 1}$ is the Bose-Einstein distribution at temperature $T$ and zero chemical potential. These equations lead to the magnon correlation function 
\begin{equation}
	 \left\langle \hat{c}_{\boldsymbol{\alpha}}^{\dagger}(\omega_1)\hat{c}_{\boldsymbol{\beta}}(\omega_2)\right\rangle = 4\pi\delta(\omega_1 + \omega_2)\delta_{\boldsymbol{\alpha}\boldsymbol{\beta}} n_{\boldsymbol{\alpha}}\operatorname{Re}\chi_{\boldsymbol{\alpha}}(\omega_2),\label{Eq:cdc} 
\end{equation} 
where the susceptibility $\chi_{\boldsymbol{\alpha}}(\omega) = \left[ - i(\omega - \omega_{\boldsymbol{\alpha}}) + \bar{\kappa}_{\boldsymbol{\mathbf{\alpha}}}/2\right]^{- 1}$ is defined analogous to Eq. (\ref{Def:chi}), and $\text{Re}\left[ \ast\right] $ denotes the real part of the argument. Similarly, 
\begin{equation}
	 \left\langle \hat{c}_{\boldsymbol{\beta}}(\omega_1)\hat{c}_{\boldsymbol{\alpha}}^{\dagger}(\omega_2)\right\rangle = \frac{n_{\boldsymbol{\alpha}} + 1}{n_{\boldsymbol{\alpha}}}\left\langle \hat{c}_{\boldsymbol{\alpha}}^{\dagger}(\omega_2)\hat{c}_{\boldsymbol{\beta}}(\omega_1)\right\rangle . \label{Eq:ccd} 
\end{equation} 
$\left\langle \hat{c}_{\boldsymbol{\alpha}}^{\dagger}(\omega_1)\hat{c}_{\boldsymbol{\beta}}(\omega_2) + \hat{c}_{\boldsymbol{\beta}}^{\dagger}( - \omega_2)\hat{c}_{\boldsymbol{\alpha}}( - \omega_1)\right\rangle $ is consistent with the fluctuation dissipation theorem \cite{LL-Stat}. By inserting different magnon distribution, $n_{\boldsymbol{\alpha}}$, we can phenomenologically deal with non-thermal cases, such as magnons excited by spin pumping or ferromagnetic resonance (FMR).\textit{\ } 

Since transmission and reflection involves different magnons, $\hat{A}_{\mathrm{T}}$ and $\hat{A}_{\mathrm{R}}$ are uncorrelated. The output power [see Eqs. (\ref{Def:Pow}) and (\ref{Aout:Full})] can therefore be written as the sum $P_{\mathrm{out}} = P_{\mathrm{el}} + P_{\mathrm{T}} + P_{\mathrm{R}}$. The purely dielectric/plasmonic contribution 
\begin{equation}
	 P_{\mathrm{el}}^{\varsigma^{\prime}}(\omega) = \sum_{\varsigma}\left\vert \delta_{\varsigma\varsigma^{\prime}} - \sum_{\mathbf{q}}\chi_{\mathbf{q}} (\omega)\sqrt{\kappa_{\boldsymbol{\mathbf{q}}}^{\varsigma^{\prime}} \kappa_{\boldsymbol{\mathbf{q}}}^{\varsigma}}\right\vert^2P_{\mathrm{in}}^{\varsigma}(\omega),\label{Pel:Gen} 
\end{equation} 
persists when $G^{\pm}\rightarrow0$. The magnonic contribution to the transmitted spectrum is 
\begin{align}
	 P_{\mathrm{T}}^{\varsigma^{\prime}}(\omega^{\prime}) &= \sum_{\varsigma \boldsymbol{\alpha}\in\mathrm{SL}}\int\frac{d\omega}{2\pi}P_{\mathrm{in}}^{\varsigma}(\omega) \nonumber \\
	 & \left[ \frac{\bar{\kappa}_{\boldsymbol{\mathbf{\alpha}}} S_{\boldsymbol{\alpha}}^+n_{\boldsymbol{\alpha}}}{\Delta_+^2 + \bar{\kappa}_{\boldsymbol{\mathbf{\alpha}}}^2/4} + \frac{\bar{\kappa}_{\boldsymbol{\mathbf{\alpha}}}S_{\boldsymbol{\alpha}}^- (n_{\boldsymbol{\alpha}} + 1)}{\Delta_-^2 + \bar{\kappa}_{\boldsymbol{\mathbf{\alpha}}}^2/4}\right] ,\label{PT:Gen} 
\end{align} 
where 
\begin{equation}
	 S_{\boldsymbol{\alpha}}^{\pm} = \left\vert \sum_{\mathbf{pq}} G_{\boldsymbol{\mathbf{pq\alpha}}}^{\pm}\sqrt{\kappa_{\boldsymbol{\mathbf{p}}}^{\varsigma}\kappa_{\boldsymbol{\mathbf{q}}}^{\varsigma^{\prime}}} \chi_{\mathbf{p}}(\omega)\chi_{\mathbf{q}}(\omega^{\prime})\right\vert^2, 
\end{equation} 
and $\Delta_{\pm} = \omega^{\prime} - \omega\mp\omega_{\boldsymbol{\alpha}}$ is the detuning from the resonance condition. Eq. (\ref{PT:Gen}) holds also after replacing $P_{\mathrm{T}}\rightarrow P_{\mathrm{R}}$ and $\mathrm{SL} \rightarrow\mathrm{LL}$. These results are general under weak coupling of any magnet to an evanescent single-mode coupler and large detuning of magnon and photon frequencies. In order to arrive at results that can be compared with experiments, we have to model $G^{\pm},$ which is done in the following. 

%%%%%Model

\section{Model} \label{Sec:Model}

The interaction of a ferromagnet interacting with light \cite{Walker57,Fletcher59,Chiasera10,Wettling75,BorovikSinha} can be described combining Maxwell's equations 
\begin{align}
	 \boldsymbol{\nabla}\times\mathbf{E} &= \frac{- \partial\mathbf{B}}{\partial t} ,\ \boldsymbol{\nabla}\times\mathbf{H} = \frac{\partial\mathbf{D}}{\partial t},\\
	 \boldsymbol{\nabla}\cdot\mathbf{B} &= 0,\ \boldsymbol{\nabla}\cdot\mathbf{D} = 0, 
\end{align} 
with the Landau-Lifshitz (LL) equation 
\begin{equation}
	 \frac{\partial\mathbf{M}}{\partial t} = - \gamma\mathbf{M}\times\mathbf{B} . 
\end{equation} 
Here $\mathbf{M}$ is the magnetization with $\left\vert \mathbf{M} \right\vert = M_s$, $\gamma$ is the absolute value of the gyromagnetic ratio, and $\mu_0$ is the vacuum magnetic permeability. The total magnetic field is $\mathbf{B} = B_0\hat{\mathbf{z}} + \mathbf{B}_L + \mathbf{B}_d$ where $B_0$ is a dc applied field that saturates the magnetization, $\mathbf{B}_L$ is the ac contribution due to light at optical frequencies, and $\mathbf{B}_d$ is the dipolar field generated by the magnetization. The magnetizing field $\mu_0\mathbf{H} = \mathbf{B} - \mu_0\mathbf{M}$, where $\mu_0$ is the vacuum magnetic permeability \cite{Borovik82,Prokhorov84}. The interaction between the magnetization and photons is modeled by a magnetization dependent permittivity in the displacement field $D_i = \epsilon_{ij}(\mathbf{M})E_j$ with $i,j\in\{x,y,z\}$, where $\epsilon_{ij}$ are the components of the permittivity tensor. \ Under weak excitation $\left\vert M_{x/y}\right\vert \ll\left\vert M_z\right\vert \approx M_s$ and $\mathbf{M} \approx M_s \hat{\mathbf{z}} + M_x\hat{\mathbf{x}} + M_y\hat{\mathbf{y}} . $ For magnons that interact with light the exchange interaction may be disregarded, since $\lambda\gg\sqrt{D/\omega}$ \cite{Walker57} where $D$ is the exchange stiffness and $\lambda$ ($\omega$) is a typical wavelength (frequency) of the magnons. This is valid for YIG with $\lambda>1\mathrm{\mu m}$ and $\omega>1$ GHz. 

The above equations can be equivalently written in terms of the Hamiltonian $H = \int d\mathbf{r}\mathcal{H}(\mathbf{r},t)$ with density \cite{Walker57,Borovik82,Demo_Tsymbal,Tianyu16} 
\begin{equation}
	 \mathcal{H} = \sum_{ij}\frac{\epsilon_{ij}(\mathbf{M})}{2}E_iE_j^{\ast} + \frac{|\mathbf{B}|^2}{2\mu_0} - \gamma\mathbf{M}\cdot\mathbf{B} . \label{GenHam} 
\end{equation} 
where $\epsilon_{ij}(\mathbf{M})$ is the \textit{permittivity tensor} to be described now. We address here a cubic material with $\{\hat{\mathbf{x}} ,\hat{\mathbf{y}},\hat{\mathbf{z}}\}$ symmetry axes such as YIG. Weak MO effects are well described by expanding the dielectric permittivity tensor $\overleftrightarrow{\epsilon}$ up to second order in the magnetization as $\overleftrightarrow{\epsilon} = \overleftrightarrow{\epsilon}^{\mathrm{el}}\left( M_s\right) + \overleftrightarrow{\epsilon}^{\mathrm{in}}\left( \mathbf{M}\right) $ \cite{Wettling75,BorovikSinha,Tianyu16}. Here 
\begin{equation}
	 \overleftrightarrow{\epsilon}^{\mathrm{el}} = 
\begin{pmatrix}
	 \epsilon_{\mathrm{s}} & - ifM_s & 0\\
	 ifM_s & \epsilon_{\mathrm{s}} & 0\\ 0 & 0 & \epsilon_{\mathrm{s}} + g^{\prime}M_s^2 
\end{pmatrix} ,\label{eps:sta} 
\end{equation} 
is called elastic because it does not lead to energy exchange between the magnetization and light. The second, \textquotedblleft inelastic\textquotedblright\ term reads 
\begin{equation}
	 \overleftrightarrow{\epsilon}^{\mathrm{in}} = 
\begin{pmatrix}
	 0 & 0 & \epsilon_{xz}\\
	 0 & 0 & \epsilon_{yz}\\ \epsilon_{xz}^{\ast} & \epsilon_{yz}^{\ast} & 0 
\end{pmatrix} ,\label{eps:dyn} 
\end{equation} 
where $\epsilon_{xz} = ifM_y + gM_sM_x$ and $\epsilon_{yz} = - ifM_x + gM_sM_y$ describe the interaction between the magnetization dynamics and the electric field. $\epsilon_{\mathrm{s}}$ is the isotropic permittivity for zero magnetization. The phenomenological constants $\{f,g,g^{\prime}\}$ parameterize the MO effects \cite{Wettling75,WettlingData76,BorovikSinha} and can be obtained directly by experiments, as we discuss now. In a magnetic material, linearly polarized light with wave vector $\mathbf{k}\parallel \mathbf{M}$ undergoes Faraday rotation defined as $\Psi_{CB}$, the rotation angle per unit length of the polarization vector. When $\mathbf{k} \perp\mathbf{M}$, linearly polarized light becomes elliptically polarized with eccentricity per unit length $\Psi_{\mathrm{LB}}^{(1)}$ for $\mathbf{M} \parallel\lbrack001]$ and $\Psi_{\mathrm{LB}}^{(2)}$ for $\mathbf{M} \parallel\lbrack111]$ (the subscripts CB and LB stand for circular and linear birefringence respectively): 
\begin{align}
	 \Psi_{\mathrm{CB}} & \approx \frac{\pi M_sf}{n_{\mathrm{s}}\epsilon_0\lambda_0},\label{CB}\\
	 \Psi_{\mathrm{LB}}^{(1)} & \approx \frac{\pi M_s^2g^{\prime}}{n_{\mathrm{s}}\epsilon_0\lambda_0},\label{LB1}\\ \Psi_{\mathrm{LB}}^{(2)} & \approx \frac{\pi M_s^2g}{n_{\mathrm{s}}\epsilon_0\lambda_0},\label{LB2} 
\end{align} 
where $n_{\mathrm{s}} = \sqrt{\epsilon_{\mathrm{s}}/\epsilon_0}$ is the refractive index, and $\lambda_0$ is the vacuum wavelength of the light. These angles suffice to fix the material parameters $\{f,g,g^{\prime}\}$. 

Circular and linear dichroism induced by absorption is negligible at frequencies below the fundamental band gap of dielectrics. For YIG this is the case when $\lambda_0>1\,\mathrm{\mu m}$ ($\omega_0<300\,$THz) \cite{Scott74}. Experiments in the configuration considered here were conducted with wave lengths $\lambda_0\sim1 . 3 - 1 . 5\,\mathrm{\mu m}$ \cite{Haigh15,ZhangWGM16,Osada16,HaighWGM}. 

The classical Hamiltonian, Eq. (\ref{GenHam}), consists of three parts $H \approx H_{\mathrm{opt}} + H_{\mathrm{mag}} + H_{\mathrm{OM}}$ and can be quantized in order to parameterize the Hamiltonian of the IO-formalism, Eq. (\ref{SQHam}). The optical part 
\begin{equation}
	 H_{\mathrm{opt}} = \int d\mathbf{r}\left[ \sum_{ij}\frac{\epsilon_{ij}^{\mathrm{el}}(M_s)}{2}E_iE_j^{\ast} + \frac{1}{2\mu_0}|\mathbf{B}_L|^2 \right] ,\label{OptHam} 
\end{equation} 
governs the normal modes of the EM fields in the presence of a static magnetization. In the magnetostatic approximation, viz. ignoring photon propagation ($c\rightarrow\infty$), and ignoring exchange, the magnetic subsystem can be described by the Hamiltonian 
\begin{equation}
	 H_{\mathrm{mag}} = \int d\mathbf{r}\left[ - \gamma\mathbf{M}\cdot\left( B_0\hat{\mathbf{z}} + \mathbf{B}_d\right) + \frac{|\mathbf{B}_d|^2}{2\mu_0} \right] ,\label{MagHam} 
\end{equation} 
as long as the samples are not too large, i.e. $a\ll c/\omega\sim1$ cm, where $a$ is the radius of the sphere \cite{Hurben95}. The optomagnonic interaction is given by 
\begin{equation}
	 H_{\mathrm{OM}} = \sum_{ij}\int d\mathbf{r}\frac{\epsilon_{ij}^{\mathrm{in}}(\mathbf{M})}{2}E_iE_j^{\ast} . \label{IntHam} 
\end{equation} 

The quantized form Eq. (\ref{SQHam}) of this classical Hamiltonian contains the matrix elements that govern the optomagnonic scattering problem that are derived in the following. 

\subsection{Whispering gallery modes} \label{Sec:WGM}

The diagonalization of Eq. (\ref{OptHam}) is equivalent to solving Maxwell's equations. We review the solutions with emphasis on WGMs [see Refs. \cite{Johnson93} and \cite{SoykalPRB10} for further details]. Let the solutions be $\mathbf{E}_{\mathbf{p}}$ and $\mathbf{B}_{\mathbf{p}}$ where $\mathbf{p}$ is a labeling of modes (to be discussed below). We expand the fields in terms of photon operators, 
\begin{equation}
	 \mathbf{\hat{E}}(\mathbf{r}) = \sum_{\mathbf{p}}\left( \mathbf{E}_{\mathbf{p}}( \mathbf{r})\hat{a}_{\mathbf{p}} + \mathbf{E}_{\mathbf{p}}^{\ast} (\mathbf{r}) \hat{a}_{\mathbf{p}}^{\dagger}\right) ,\label{Elexp} 
\end{equation} 
where $\hat{a}_{\mathbf{p}}$ is the annihilation operator for $\mathbf{p}$ -mode. A similar equation holds with $\hat{\mathbf{E}}\rightarrow \hat{\mathbf{B}}_L$ and $\mathbf{E}_{\mathbf{p}}\rightarrow\mathbf{B}_{\mathbf{p}}$. Inserting these into the optical Hamiltonian, Eq. (\ref{OptHam}), recovers the second-quantized Hamiltonian, $\hat{H}_{\mathrm{opt}} = \sum\hbar\omega_{\mathbf{p}}\hat{a}_{\mathbf{p}}^{\dagger}\hat{a}_{\mathbf{p}}$, after normalizing the field amplitudes as explained in Appendix \ref{App:ScatAmp}. 

The WGMs of large spheres are linearly polarized plane waves moving adiabatically along the equator with transverse electric (TE, $\mathbf{E} \parallel\hat{\mathbf{z}}$) or transverse magnetic (TM, $\mathbf{E} \parallel\hat{\mathbf{r}}$) polarization. TE and TM modes are degenerate in axially symmetric wires but not in a sphere \cite{Schiller91}. This degeneracy is broken by the surface, which is known as geometrical birefringence \cite{Lan11}. When $\mathbf{M} = 0$, the angular momentum of WGMs in spheres is conserved. When $\mathbf{M}\neq0$ a conserved angular momentum is still a good approximation since MO effects are weak \cite{Haigh15}. 

The optical modes in systems with spherical symmetry are fully described by the collective index $\mathbf{p}\equiv\{\nu,l,m,\sigma\},$ where $\sigma \in\{\mathrm{TE},\mathrm{TM}\}$ is the polarization index and $\{\nu,l,m\}$ are integers satisfying $\nu,l>0$ and $\left\vert m\right\vert \leq l$ \cite{Chiasera10}. The total ($z$-component of the) angular momentum $\mathbf{L}$ of a mode $\mathbf{p}$ is $\left\vert \mathbf{L}\right\vert = \hbar l$ ($L_z = \hbar m$). $\nu - 1$ is the number of nodes of the electric field amplitude in the radial direction. WGMs are those modes which satisfy $l\gg1$ and $\left\vert m\right\vert /l \approx 1$. The sign of $m$ governs their circulation direction; $m>0$ and $m<0$ refers to blue and green arrows respectively in Fig. \ref{Setup}. 

The field in the sphere is distributed as \cite{Johnson93} 
\begin{align}
	 \mathbf{E}_{\mathbf{p}(\mathrm{TE})} &= \mathcal{E}_{\mathbf{p}} j_l(k_{\mathbf{p}}r)\mathbf{Y}_l^m(\theta,\phi),\label{ETE}\\
	 \mathbf{E}_{\mathbf{p}(\mathrm{TM})} &= \frac{\mathcal{E}_{\mathbf{p}}}{k_{\mathbf{p}}}\boldsymbol{\nabla}\times\left[ j_l(k_{\mathbf{p}}r)\mathbf{Y}_l^m(\theta,\phi)\right] ,\label{ETM} 
\end{align} 
where $\mathcal{E}_{\mathbf{p}}$ is the normalization constant derived in Appendix \ref{App:ScatAmp}. $j_l$ is the spherical Bessel function of the first kind. $\mathbf{Y}_l^m = \boldsymbol{\mathcal{L}}Y_l^m/\sqrt{l(l + 1)}$ is a vector spherical harmonic generated by operating with the dimensionless angular momentum $\boldsymbol{\mathcal{L}} = - i\mathbf{r} \times\boldsymbol{\nabla}$ on the scalar spherical harmonic $Y_l^m$ [see Eq. (\ref{Exp:Ylm})]. $k_{\mathbf{p}}$ can be interpreted as the wave vector related to the frequency by $\omega_{\mathbf{p}} = ck_{\mathbf{p}}/n_{\sigma}$ , and does not depend of $m$. The dispersion relation for the WGMs \cite{Johnson93,Haigh15} 
\begin{equation}
	 k_{\mathbf{p}}a = \omega_{\mathbf{p}}\frac{an_{\sigma}}{c} = l + \beta_{\nu}\left( \frac{l}{2}\right)^{1/3} - P_{\sigma} + O\left( l^{- 1/3}\right) ,\label{ResWGM} 
\end{equation} 
where $\beta_{\nu}>0$ is the $\nu$-th root of the Airy function on the negative real axis, $P_{\mathrm{TE}} = n_{\mathrm{s}}/\sqrt{n_{\mathrm{s}}^2 - 1}$, $P_{\mathrm{TM}}^{- 1} = n_{\mathrm{s}}\sqrt{n_{\mathrm{s}}^2 - 1}$, and $a$ is the sphere radius. The refractive indices differ for the two polarizations, with $n_{\mathrm{TM}} = n_{\mathrm{s}}$ and 
\begin{equation}
	 n_{\mathrm{TE}} = n_{\mathrm{s}}\left( 1 + \frac{\Psi_{LB}^{(1)}}{k_p}\right) . 
\end{equation} 
The frequency difference at constant $l$ and $\nu$ for the two polarizations is 
\begin{equation}
	 \omega_{\mathrm{TM}} - \omega_{\mathrm{TE}} = \frac{c}{n_{\mathrm{s}}}\left( \frac{\sqrt{n_{\mathrm{s}}^2 - 1}}{an_{\mathrm{s}}} + \Psi_{LB}^{(1)}\right) ,\label{LinBireF:Orig} 
\end{equation} 
with contributions from both geometric and magnetic linear birefringence \cite{Haigh15}. 

We now turn to the spatial amplitude distribution of the WGMs in polar coordinates $\{r,\theta,\phi\}$. For $l\gg1$ Eqs. (\ref{ETE},\ref{ETM}) become 
\begin{align}
	 \mathbf{E}_{\mathbf{p}(\mathrm{TE})} & \propto j_l(k_{\mathbf{p}} r)P_l^m(\cos\theta)e^{im\phi}\hat{\mathbf{z}},\\
	 \mathbf{E}_{\mathbf{p}(\mathrm{TM})} & \propto j_l(k_{\mathbf{p}} r)P_l^m(\cos\theta)e^{im\phi}\hat{\mathbf{r}}, 
\end{align} 
where $P_l^m$ is the associated Legendre polynomial of degree $l$ and order $m$. $e^{im\phi}\hat{\mathbf{z}}$ can be interpreted as a $\hat{\mathbf{z}}$ polarized wave traveling around the azimuthal. The sign of $m$ decides the chirality of WGM with $m>0$ corresponding to angular momentum along $ + \hat{\mathbf{z}}$ (blue arrow in Fig \ref{Setup}). 

\begin{figure}[ptb]
\begin{equation*}
\includegraphics[width=0.48\textwidth,keepaspectratio]{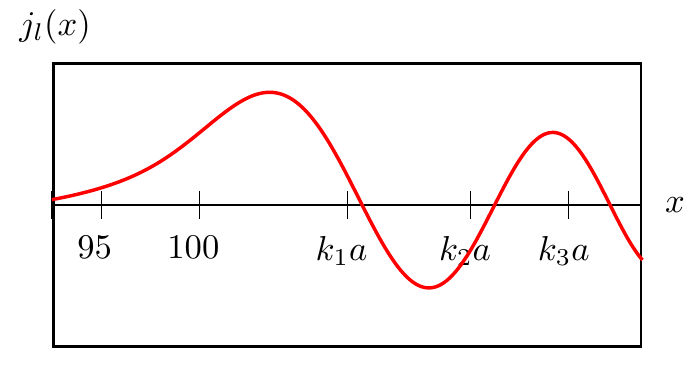}
\end{equation*}
\caption{Spherical Bessel function of the first kind $j_{l}(x)$ for $l=100$
and $x\approx l$. For a given WGM with mode $\mathbf{p}$, the above graph can
be converted from $x$ to the radial coordinate by $r=x/k_{\mathbf{p}}$ with
$r\leq a$ lying inside the sphere. $k_{\mathbf{p}}$ is given by Eq. (
\ref{ResWGM}) with $\mathbf{p}=\{\nu,100,m,\mathrm{TE}\}$, independent of $m$
and on the scale of the graph also on the polarization: $k_{1}a=108.6$ ,
$k_{2}a=115.1$, and $k_{3}a=120.4$. }
\label{Fig:Bessel}
\end{figure}

The Bessel function $j_l(x)$ is plotted in Fig. \ref{Fig:Bessel} for $l = 100 $ ($j_l(k_{\mathbf{p}}r)$ for other $l\gg1$ are analogous). For $x\ll l$, $j_l(x) \approx 0$. For $x\sim l$, it is oscillatory with zeroes at $x_{\nu} \approx l + \beta_{\nu}(l/2)^{1/3}$, where $\beta_{\nu} \approx 1 . 5 + 1 . 2\nu$ are the zeroes of Airy's function for $\nu\sim1 - 10$. For a WGM with label $\mathbf{p}$, $x\leq k_{\mathbf{p}}a$ corresponds to the amplitude inside the sphere. The modes with $\nu = 1,2,3$ reach the sphere surface at the valued marked in Fig. \ref{Fig:Bessel}, showing that there are $\nu - 1$ radial nodes in this interval. The electric fields for higher $\nu$ are less localized. Weaker electric fields at the surface lead to weaker proximity coupling to the optical fiber or prism. 

Close to the equator the Legendre polynomial for $l = m\gg1$ is a Gaussian function 
\begin{equation}
	 P_l^l(\theta)\propto\sin^l\theta \approx \exp\left[ - \frac{l}{2} \left( \theta - \frac{\pi}{2}\right)^2\right] , 
\end{equation} 
with width $1/\sqrt{l}$ centered at $\theta = \pi/2$, reflecting the confinement of the WGMs to the equatorial plane. For $\left\vert m\right\vert \neq l$ and $1 - \left\vert m\right\vert /l\ll1$, $P_l^m(\theta) \approx p_l^m(\theta)P_m^m(\theta),$ where $p_l^m(\theta)$ is a polynomial in $\theta$ of degree $l - m$. Therefore, $P_l^m(\theta)$ has $l - m$ roots and decays like a Gaussian with a length scale $1/\sqrt{\left\vert m\right\vert}$. This implies that for a fixed $l$ , the WGMs couple weaker to the optical fiber with decreasing $\left\vert m\right\vert <l$. 

\subsection{Magnetostatic modes} \label{Sec:Magnons}

Here we consider the spin waves that diagonalize the magnetic Hamiltonian Eq. (\ref{MagHam}) for spherical magnets with equilibrium magnetization along $\mathbf{\hat{z} .}$ The eigenstates are fully characterized by three integers $\boldsymbol{\alpha} = \{\nu_s\geq0,l_s>0,m_s\},$ $\left\vert m_s\right\vert \leq l_s$. The $z$-component of the total angular momentum $L_z = \hbar m_s$ \cite{Walker57,Fletcher59}, while $\hbar l_s$ can be interpreted as its total angular momentum. The index $\nu_s$ can be associated with the number of nodes in the radial amplitude function \cite{Walker57,Fletcher59}. Note that $\nu_s\geq0$ \cite{Walker57}, unlike $\nu>0$ for WGMs. In second quantized notation, we can expand the magnetization field as \cite{AkashSN16} 
\begin{equation}
	 \hat{M}_-(\mathbf{r}) = \sum_{\boldsymbol{\alpha}}\left[ M_{\boldsymbol{\alpha}, -}(\mathbf{r})\hat{c}_{\boldsymbol{\alpha}} + \left( M_{\boldsymbol{\alpha} , +}(\mathbf{r})\right)^{\ast}\hat{c}_{\boldsymbol{\alpha}}^{\dagger}\right] ,\label{Magexp} 
\end{equation} 
where $M_{\boldsymbol{\boldsymbol{\alpha}},\pm} = \mathbf{M}_{\boldsymbol{\alpha}}\cdot\left( \hat{\mathbf{x}}\pm i\hat{\mathbf{y}}\right) $, $\mathbf{M}_{\boldsymbol{\alpha}}$ is the amplitude of the $\boldsymbol{\alpha}$ -magnon and $\hat{c}_{\boldsymbol{\alpha}}$ is its annihilation operator. Inserting this expansion into Eq. (\ref{MagHam}) leads to the Hamiltonian $\hat{H}_{\mathrm{mag}} = \sum_{\boldsymbol{\alpha}} \hbar\omega_{\boldsymbol{\alpha}}\hat{c}_{\boldsymbol{\alpha}}^{\dagger} \hat{c}_{\boldsymbol{\alpha}}, $ where the magnetization profiles are normalized as described in Appendix \ref{App:ScatAmp}. While the full expressions for $\mathbf{M}_{\boldsymbol{\alpha}}$ are complicated \cite{Walker57,Fletcher59}, they become manageable for the magnons that contribute significantly to the scattering of optical WGMs, viz. in the long wavelength limit with $\left\vert m_s\right\vert \sim1$ and the surface modes with $\left\vert m_s\right\vert \gg1$. 

\emph{Small-L:} The magnons with $l_s,\left\vert m_s\right\vert \sim1$ extend through the whole sphere with wavelengths $\lambda\sim a$. The Kittel mode $\mathbf{u} = \{0,1,1\}$ is a uniform precession with $M_{\mathbf{u}, -}( \mathbf{r}) = M_u$ and $M_+ = 0$, where $M_u = \sqrt{4\gamma\hbar M_s/V},$ see Eq. (\ref{Val:Mu}) in Appendix \ref{App:ScatAmp}. This magnon has a (purely) spin angular momentum $\hbar$. The resonant frequency $\omega_{\mathbf{u}} = \gamma(B_0 - \mu_0M_s)$. 

Explicit expressions for the spatial profiles of finite but small angular momentum magnons are given in \cite{Fletcher59} for spheres with (free) boundary conditions from the Maxwell's equations. In polar coordinates $\{r,\theta,\phi\}$ and near the equator, 
\begin{equation}
	 M_{\boldsymbol{\alpha},\pm} \approx \mathcal{M}_{\pm}Y_{l_s}^{m_s}(\theta, \phi)\frac{e^{\pm i\phi}}{\sin\theta}, 
\end{equation} 
where $\mathcal{M}_{\pm}$ are constants with $\left\vert \mathcal{M}_+\right\vert \ll\left\vert \mathcal{M}_-\right\vert $ and $\left\vert \mathcal{M}_-\right\vert \sim M_u$. $\mathcal{M}_{\pm}$ depends only weakly on $\boldsymbol{\alpha}$. We are not aware of a general formula for the resonant frequencies $\omega_{\boldsymbol{\alpha}}$. However, they depend on both $m_s$ and $l_s$ and lie in the interval 
\begin{equation*}
	 B_0 - \frac{4\mu_0M_s}{3}\leq\frac{\omega_{\boldsymbol{\alpha}}}{\gamma} \leq B_0 - \frac{5\mu_0M_s}{6} . 
\end{equation*} 
They do not depend on the sphere radius. For the special case of $m_s = l_s $, 
\begin{equation}
	 \omega_{\boldsymbol{\alpha}} = \gamma B_0 - \frac{5l_s + 4}{3(2l_s + 1)} \gamma\mu_0M_s, 
\end{equation} 
while for $m_s = l_s - 1$, 
\begin{equation}
	 \omega_{\boldsymbol{\alpha}} = \gamma B_0 - \frac{5l_s + 7}{3(2l_s + 1)} \gamma\mu_0M_s . \label{OmegaMag:lm1} 
\end{equation} 
Both the cases have only one mode, which is nodeless in the radial direction, labeled as $\nu_s = 0$. Their frequencies lie both above and below the Kittel mode. For instances, $\omega_{\left\{0,2,2\right\}} - \omega_{\mathbf{u}} = \gamma\mu_0M_s/15$ and $\omega_{\left\{0,2,1\right\}} - \omega_{\mathbf{u}} = - 2\gamma\mu_0M_s/15$. For larger $l - m $, the eigenfrequencies are the solutions of polynomial equations listed in Table VII of \cite{Fletcher59} for a small number of $\nu_s$. 

\emph{Large-L:} The magnons with large angular momenta $m_s\gg1$ are chiral Damon-Eshbach (DE) modes localized at the equator \cite{DamEshSurface,Fletcher59}. These are magnetic analogues to the whispering gallery modes when $l_s \approx m_s$, i.e. spatially confined to the equator. A magnon with index $\boldsymbol{\mathcal{D}} = \{0,l_s ,l_s\}$ and $l_s\gg1$ has amplitude $M_+ = 0$ and \cite{Fletcher59} 
\begin{equation}
	 M_{\boldsymbol{\mathcal{D}}, -}(\mathbf{r}) = M_{l_s}\left( \frac{r\sin\theta}{a}\right)^{l_s - 1}e^{i(l_s - 1)\phi},\label{DE:Dist} 
\end{equation} 
where $M_{l_s} = \left( l_s/\pi\right)^{3/4}\sqrt{4\gamma\hbar M_s/a^3}$ , see Eq. (\ref{Val:Mls}). We can interpret it as a plane wave running counterclockwise (blue arrow in Fig. \ref{Setup}) along the equator with (local) linear momentum $k_{\boldsymbol{\mathcal{D}}}\boldsymbol{\hat{\phi}}$ with $k_{\boldsymbol{\mathcal{D}}} \approx a/l_s$. This corresponds to a circular motion with positive orbital angular momentum $L_{\boldsymbol{\mathcal{D}}}/\hbar = k_{\boldsymbol{\mathcal{D}}}a \approx l_s\gg1$. The DE magnetization decays exponentially as a function of distance from the interface with a length scale $a/l_s$, i.e. the same as the azimuthal wavelength $k_{\boldsymbol{\mathcal{D}}}^{- 1}$. The strict confinement of DE modes is in contrast to the WGMs that decay more slowly as shown in Fig. \ref{Fig:Bessel}. For large $k_{\boldsymbol{\mathcal{D}}}$ the DE modes are degenerate at $\omega_{\mathrm{DE}} = \gamma\left( B_0 - 5\mu_0M_s/6\right) , $ i.e. blue shifted with respect to the Kittel modes. As for the small-$L$ magnons, the amplitude of the mode with $l_s = m_s$ has no radial nodes and $\nu_s = 0$. 

The amplitudes of DE modes with $l_s\neq m_s$ are complicated \cite{Fletcher59} but qualitatively similar to the above form as long as $l_s - m_s\ll l_s$. We derive below that these magnons reflect photons with momentum $k_{\mathbf{p}}$ when the condition $k_{\boldsymbol{\mathcal{D}}} \approx 2k_{\mathbf{p}}$ is fulfilled and contribute to BLS with roughly equal scattering amplitudes. 

\section{BLS Amplitude} \label{Sec:Coup}

Here we calculate the coupling between the WGMs and magnons as expressed by Eq. (\ref{IntHam}) by perturbation theory. Inserting the magnon and photon non-interacting normal modes derived above [see Eqs. (\ref{Elexp}) and (\ref{Magexp})] into Eq. (\ref{IntHam}), we find that the interaction Hamiltonian reduces to the form considered previously, Eq. (\ref{Hint}), with coupling constants written below. 

The absence of diagonal terms in Eq. (\ref{eps:dyn}) implies $G_{\sigma = \sigma^{\prime}}^{\pm} = 0$. In other words, $\mathrm{TE}\rightarrow \mathrm{TE}$ and $\mathrm{TM}\rightarrow\mathrm{TM}$ scattering probability vanishes, implying that the incident and the scattered photons have orthogonal polarizations. Let $\mathbf{p}\equiv\{\nu,l,m,\mathrm{TE}\}$, $\mathbf{q} \equiv\{\nu^{\prime},l^{\prime},m^{\prime},\mathrm{TM}\}$, and $\boldsymbol{\alpha}\equiv\{\nu_s,l_s,m_s\}$ as in the previous section to arrive at 
\begin{align}
	 \hbar G_{\mathbf{pq}\boldsymbol{\alpha}}^+ &= \frac{\mathcal{G}_+}{4}\int E_{\mathbf{p},z}\left( E_{\mathbf{q}, +}^{\ast} M_{\boldsymbol{\boldsymbol{\alpha}}, -} + E_{\mathbf{q}, -}^{\ast} M_{\boldsymbol{\alpha}, +}\right) d\mathbf{r},\label{Def:Gp}\\
	 \hbar G_{\mathbf{pq}\boldsymbol{\alpha}}^- &= \frac{\mathcal{G}_-}{4}\int E_{\mathbf{p},z}\left( E_{\mathbf{q}, +}^{\ast} M_{\boldsymbol{\boldsymbol{\alpha}}, -}^{\ast} + E_{\mathbf{q}, -}^{\ast}M_{\boldsymbol{\alpha}, +}^{\ast}\right) d\mathbf{r},\label{Def:Gm} 
\end{align} 
where $E_{\mathbf{q},\pm}^{\ast} = \left( \mathbf{E}_{\mathbf{q}}^{\ast}\right) \cdot\left( \hat{\mathbf{x}}\pm i\hat{\mathbf{y}}\right) $ etc. and $\mathcal{G}_{\pm} = gM_s\pm f$. 

We have four possible incident WGMs, with $\sigma\in\{\mathrm{TE} ,\mathrm{TM}\}$ and $m \approx \pm l$. In the following we explicitly illustrate the concepts for the particular case of TE polarized incident WGM with $m>0$ (rotation sense of blue arrow in Fig. \ref{Setup}). Subsequently, we give the results for $m<0$, while the case of TM polarized input follow from Hermiticity, $G_{\mathbf{qp}\boldsymbol{\alpha}}^{\pm} = \left( G_{\mathbf{pq} \boldsymbol{\alpha}}^{\mp}\right)^{\ast}$. 

\subsection{Small-$L$} \label{Coup:SL}

For $m>0$, the integrals, Eqs. (\ref{Def:Gp}) and (\ref{Def:Gm}), can be simplified for the Kittel mode (see Appendix \ref{App:KittScat}), 
\begin{equation}
	 G_{\mathbf{pq}\mathbf{u}}^{\pm} = g_{\pm}\delta_{\nu,\nu^{\prime}} \delta_{m^{\prime},m\pm1}\delta_{l - m,l^{\prime} - m^{\prime}} ,\label{Coup:Kitt} 
\end{equation} 
where 
\begin{equation}
	 g_{\pm} = \frac{c\left( \Psi_{LB}^{(2)}\pm\Psi_{CB}\right)}{2n_s\sqrt{sV}},\label{Def:gpm} 
\end{equation} 
with $s = M_s/(\gamma\hbar)$ the spin (number) density and $V$ the volume of the sphere. 

The orthogonality of WGMs and constant amplitude of the Kittel mode leads to the selection rule $\nu = \nu^{\prime}$. The $z$-component of the total angular momentum is conserved when $m^{\prime} = m\pm1$, where the upper(lower) sign corresponds to annihilation(creation) of a magnon. The third selection rule, $l^{\prime} - m^{\prime} = l - m$ can be interpreted as the conservation of the non-$z$ component of angular momentum since $l - m\propto l^2 - m^2\propto L^2 - L_z^2$. This condition is not exact when rotational symmetry is broken by the magnetization, but a good approximation here by the smallness of the MO coupling. 

We can extend the discussion to small but finite-$L$ magnons. The coupling constant for Stokes scattering is 
\begin{align}
	 G_{\mathbf{pq}\boldsymbol{\alpha}}^- & \propto\delta_{\nu,\nu^{\prime}}\int\left( Y_{l_s}^{m_s}\right)^{\ast}Y_l^m\left( Y_{l^{\prime}}^{m^{\prime}}\right)^{\ast}d\boldsymbol{\Omega} \nonumber \\
	 & \propto\delta_{\nu,\nu^{\prime}}\left\langle l^{\prime},0;l_s ,0\right\vert \left . l,0\right\rangle \left\langle l^{\prime},m^{\prime};l_s,m_s\right\vert \left . l,m\right\rangle ,\label{LowAM} 
\end{align} 
where $d\boldsymbol{\Omega} = \sin\theta d\theta d\phi . $ The Clebsch-Gordon (CG) coefficient $\left\langle l_1,m_1;l_2,m_2\right\vert \left . l_3,m_3\right\rangle $ is the amplitude of two angular momentum states $\{l_1,m_1\}$ and $\{l_2,m_2\}$ adding up to a third $\{l_3,m_3\}$ , with explicit expressions in for instance \cite{Sakurai} . If we interpret $l_s$ as the angular momentum of a magnon, the first and second CG coefficients express conservation of $L$ and $L_z$ respectively. The coupling strengths depend on the transverse magnetization of the corresponding magnon at the equatorial surface that is of the same order as the Kittel mode, leading to the estimate, 
\begin{equation}
	 G_{\mathbf{pq}\boldsymbol{\alpha}}^-\sim g_-\delta_{\nu,\nu^{\prime}}\left\langle l^{\prime},0;l_s,0\right\vert \left . l,0\right\rangle \left\langle l^{\prime},m^{\prime};l_s,m_s\right\vert \left . l,m\right\rangle . \label{Eq:GmTra} 
\end{equation} 
Analogously, the anti-Stokes scattering is governed by 
\begin{equation}
	 G_{\mathbf{pq}\boldsymbol{\alpha}}^+\sim g_+\delta_{\nu,\nu^{\prime}}\left\langle l,0;l_s,0\right\vert \left . l^{\prime},0\right\rangle \left\langle l,m;l_s,m_s\right\vert \left . l^{\prime},m^{\prime}\right\rangle . \label{Eq:GpTra} 
\end{equation} 
When $m<0$ (rotation sense of green arrow in Fig. \ref{Setup}), a similar calculation shows that the above results are valid for negative $m,m^{\prime} $ as well. 

\subsection{Large-$L$} \label{Coup:LL}

While small angular momentum-magnons scatter light into the forward direction, light can be backscattered by magnons with angular momenta twice of that of the photon. We focus on the chiral DE magnons that encircle the equatorial surface with mode numbers $l_s = m_s\gg1$. The conservation of $L_z$ gives $m^{\prime} = m\mp l_s$ where the upper(lower) sign refers to creation(annihilation) of a magnon. As discussed in Sec. \ref{Sec:Qual}, $m \approx - m^{\prime}$ by energy conservation, and therefore the only allowed transition is with $m^{\prime} = m - l_s$ with $l_s \approx 2m$. In other words, in the present configuration a WGM can be scattered backward only by creating a magnon, but not by annihilating one. In Appendix (\ref{App:DEScat}), we derive for $m>0$ and $m^{\prime}<0$, 
\begin{equation}
	 G_{\mathbf{pq\boldsymbol{\mathcal{D}}}}^- \approx \Xi_-g_-\left\langle l,0;l^{\prime},0\right\vert \left . l_s,0\right\rangle \left\langle l,m;l^{\prime},\left\vert m^{\prime}\right\vert \right\vert \left . l_s,l_s\right\rangle ,\label{Coup:DE} 
\end{equation} 
and $G_{\mathbf{pq\boldsymbol{\mathcal{D}}}}^+ = 0$, where $g_-$ is given by Eq. (\ref{Def:gpm}). The pre-factor 
\begin{equation}
	 \Xi_- = ( - 1)^{\nu - \nu^{\prime} + m^{\prime}}\sqrt{\frac{4}{3}}\pi P_{TE}\left( 1 + P_{TM}\right) . \label{Prefm} 
\end{equation} 
is of order $\left\vert \Xi_-\right\vert \sim1 . $ There is no selection rule for the radial mode indices. The CG coefficients imply that the scattering is non-zero only when $m = l_s + m^{\prime}$ as argued above. The scattering amplitude is maximized when the angular momentum is conserved $l + l^{\prime} \approx l_s$. 

A similar calculation for a WGM with opposite circulation $m<0$ and $m^{\prime}>0$, gives $G_{\mathbf{pq\boldsymbol{\mathcal{D}}}}^- = 0$ 
\begin{equation}
	 G_{\mathbf{pq\boldsymbol{\mathcal{D}}}}^+ \approx \Xi_+g_+\left\langle l,0;l^{\prime},0\right\vert \left . l_s,0\right\rangle \left\langle l,|m|;l^{\prime},m^{\prime}\middle|l_s,l_s\right\rangle , 
\end{equation} 
with 
\begin{equation}
	 \Xi_+ = ( - 1)^{\nu - \nu^{\prime} + m^{\prime}}\sqrt{\frac{4}{3}}\pi P_{TE}\left( 1 - P_{TM}\right) . \label{Prefp} 
\end{equation}

The above coupling constants are dependent on the overlap of DE magnons and WGMs as given in Eqs. (\ref{Int:AngDE}) and (\ref{Int:RadDE}). The angular overlap gives the angular momentum conservation laws selecting the DE magnon based on $\mathbf{p}$ and $\mathbf{q}$. For given WGMs and DE magnons, the radial overlap is small owing to two factors. First, WGMs have an node close to the surface at which the DE magnon amplitude is largest [see Fig. \ref{Fig:Bessel}]. Secondly, the spatial distributions of WGMs are wider ($\sim a/l^{2/3}$) than those of the DE modes ($\sim a/l_s$). By engineering the spatial distribution of WGMs, the overlap can possibly be enhanced, as will be discussed in a forthcoming article. 

%%%%%% I/O Formalism 

\section{Transmission and reflection spectra} \label{Sec:TxRx}

With the expressions for $G^{\pm}$ in hand, we can calculate the transmitted and the reflected spectrum given any input spectrum, $P_{\mathrm{T}}$ and $P_{\mathrm{R}}$ in terms of $P_{\mathrm{in}}$ [see Eq. (\ref{PT:Gen})]. In principle, the output power spectrum can be numerically evaluated from the expressions derived above. Analytical expressions for the general case are complicated and difficult to interpret. Leaving this task for future work, we focus here on a special case to illustrate our results. The notation has been defined in Sec. \ref{Sec:Output}. 

\subsection{Setup} \label{Sec:Setup}

\textit{Coupling:} The evanescent coupling of a magnetic sphere can be achieved by proximity to an optical fiber or prism that is illuminated by photons with tunable frequency $\omega$, wave-vector $k$, and polarization $\varsigma$. We assume dominantly adiabatic coupling in which only WGMs with matching polarizations ($\sigma = \varsigma$) and wave vectors ($m \approx ka$) are populated. Under these conditions, the leakage from and to the fiber into a mode $\mathbf{w}\equiv\{\nu,l,m,\sigma\}$ is $\kappa_{\boldsymbol{\mathbf{w}}}^{\varsigma} = \delta_{\varsigma,\sigma}\kappa_{\mathbf{w}}$, where $\kappa_{\mathbf{w}}$ is a constant depending on the precise system parameters. 

The resonance condition holds for large $l\lessapprox\omega n_sa/c$, with precise value of $l$ discussed below. For a single mode fiber with a contact point to the sphere much smaller than the wavelength, the wave-vector matching holds only approximately and WGMs with many $m$-values can be excited. However, the coupling can be engineered by tapering the fiber to a width below the wavelength as discussed in \cite{Humphrey07}. This additional degree of freedom allows to match modes and selectively enhance the coupling to WGMs with small $l - |m|$ and $\nu$. Here we consider the case where $\kappa_{\mathbf{w}}$ is significant only for $\nu\in\{1,2\}$ and is maximal at $m = l$ (for a given $\{\nu,l,\sigma\}$). These assumptions can be verified in a particular experiment by monitoring the elastic transmission power $P_{\mathrm{el}}$ in Eq. (\ref{Pel:Gen}) \cite{Haigh15}. 

\textit{Sphere:} Let us consider a YIG sphere of radius $a = 200\,\mathrm{\mu m .}$ At room temperature $M_s = 1 . 4\times10^5$ A/m and $n_s = 2 . 2$ . The incident light has wavelength $\lambda_0 \approx 1\,\mathrm{\mu m}$ and is tunable. Near this wavelength, the MO constants are $\Psi_{CB} \sim500\,\text{rad/m}$ and $\Psi_{LB}^{(1)} = \Psi_{LB}^{(2)}\sim 200\,\text{rad/m}$ \cite{WettlingData76}, which leads to $g_+ = 2\pi\times6$ Hz and $g_- = - 2\pi\times2 . 6$ Hz [see Eq. (\ref{Def:gpm})]. The latter numbers agree with the estimate $g = 2\pi\times5$ Hz from \cite{Osada16}, where it is not clearly specified whether $g$ is $g_+$ or $g_-$. 

A magnetic field $B_0$ shifts the magnon frequencies rigidly by the Zeeman energy. In thermal equilibrium at room temperature with $\omega_{\boldsymbol{\alpha}}\sim1 - 10$ GHz, we have $\hbar\omega_{\boldsymbol{\alpha}}n_{\boldsymbol{\alpha}} \approx k_BT$ and $n_{\boldsymbol{\alpha}}\gg1$. When the sample is excited by resonant microwaves, the Kittel mode is selectively populated and $n_{\boldsymbol{u}}$ can become much larger than the thermal population. $\bar{\kappa}_{\boldsymbol{\mathbf{\alpha}}}\sim\alpha_G \omega_{\boldsymbol{\alpha}},$ where $\alpha_G = 10^{- 4}$ is typical for Gilbert damping in YIG \cite{WuHoff}. 

\begin{figure}[ptb]
\begin{equation*}
\includegraphics[width=.48\textwidth,keepaspectratio]{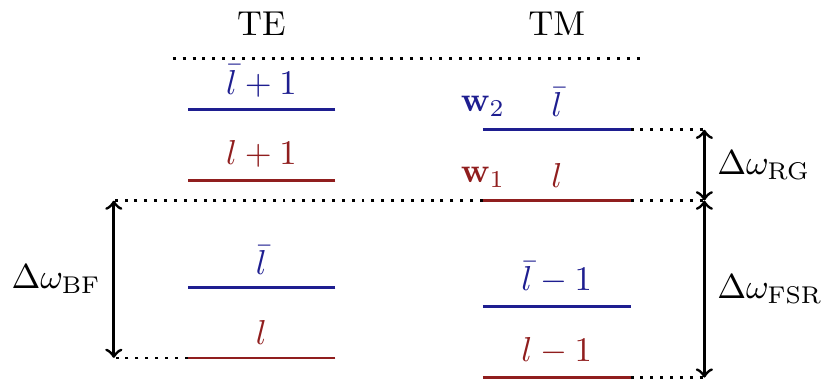}
\end{equation*}
\caption{Spectrum of WGMs: The resonant frequency of a WGM depends on the
angular momentum ($l_{R}$), the number of radial nodes ($\nu-1$), and the
polarization. This sketch includes the levels for the first two radial modes
$\nu=1$ and $\nu=2$ denoted by red and blue respectively. The labels for the
splittings ($\mathrm{FSR}$, $\mathrm{BF}$, and $\mathrm{RG}$) are defined in
the text.}
\label{WGMEnergies}
\end{figure}

\textit{WGM spectrum:} The frequencies from Eq. (\ref{ResWGM}) are sketched in Fig. \ref{WGMEnergies}. They depend on $l$, $\nu$, and $\sigma$ (but not on $m$). For fixed $l$ and $\nu$, the frequencies for two polarizations differ by $\Delta\omega_{\mathrm{BF}}$, given in Eq. (\ref{LinBireF:Orig}), 
\begin{equation}
	 \Delta\omega_{\mathrm{BF}} = \frac{c}{n_{\mathrm{s}}a}\left( \frac{\sqrt{n_{\mathrm{s}}^2 - 1}}{n_{\mathrm{s}}} + a\Psi_{LB}^{(1)}(\lambda_0)\right) . \label{LinBireF:Rep} 
\end{equation} 
The gap between $l$ and $l + 1$ is called ``free spectral range (FSR)\textquotedblright. For a fixed $\sigma$ and $\nu$, \ $\Delta \omega_{\mathrm{FSR}} \approx c/\left( n_sa\right) $ [see Eq. (\ref{ResWGM})]. Two ladders of WGMs with $\nu = 1$ and $\nu = 2$ are shown by red and blue in Fig. \ref{WGMEnergies}, respectively (we consider only two $\nu$ values as discussed before). The splitting between levels with different $\nu$ but same $l$ is large $\sim4\mathrm{THz}$, but levels can be close for different angular momenta. For a given $l$, we define $\bar{l}$ as the WGM in the $\nu = 2$ branch with frequency just above the $\{\nu = 1,l\}$ WGM. So $\bar{l}$ is the lowest integer such that $\omega_{\mathbf{w}_2}>\omega_{\mathbf{w}_1}$ where $\mathbf{w}_i$ are defined in Fig. \ref{WGMEnergies}. For large $l,\bar{l}$ 
\begin{equation}
	 \omega_{\mathbf{w}2}>\omega_{\mathbf{w}1}\Rightarrow\bar{l}>l - \frac{\beta_2 - \beta_1}{2^{1/3}}l^{1/3}, 
\end{equation} 
where $(\beta_2 - \beta_1)/2^{1/3} \approx 1 . 4$. This gives $\bar{l} = \left\lceil l - 1 . 4l^{1/3}\right\rceil $ where the ceiling function $\lceil x\rceil$ is the smallest integer greater than $x$. We define the ``radial gap\textquotedblright\ $\Delta\omega_{\mathrm{RG}} = \omega_{\mathbf{w} 2} - \omega_{\mathbf{w}1}$ that depends on the fractional part of $l - 1 . 4l^{1/3} $. The scattering between modes with different $\nu$ can be relevant in reflection, as discussed below. 

We can estimate the characteristic frequency splittings in Fig. \ref{WGMEnergies} for our model system as follows. We tune the input laser frequency $\omega_{\mathrm{in}}$ to the mode $\{\nu = 1,l_p = 1257,l_p, \mathrm{TE}\}$ (at a wavelength around $1\,\mathrm{\mu}$m). We find $\bar{l}_p = 1242$ and $\Delta\omega_{\mathrm{RG}} = 2\pi\times16$ GHz. This is much smaller than the free spectral range $\Delta\omega_{\mathrm{FSR}} = 2\pi\times108$ GHz and birefringence $\Delta\omega_{\mathrm{BF}} = 2\pi \times101$ GHz. 

\textit{Source:} Let us assume TE polarized input light (TM is discussed below) at frequency $\omega_{\mathrm{in}}$ (defined above). Its power spectrum is 
\begin{equation}
	 P_{\mathrm{in}}^{\varsigma}(\omega) = 2\pi\delta_{\mathrm{TE},\varsigma} \bar{P}_{\mathrm{in}}\delta(\omega - \omega_{\mathrm{in}}),\label{Def:Pin} 
\end{equation} 
where $\bar{P}_{\mathrm{in}}$ is the total integrated power in the input 
\begin{equation}
	 \bar{P}_{\mathrm{in}} = \int\frac{d\omega}{2\pi}P_{\mathrm{in}}^{\mathrm{TE}}(\omega) . \label{Eq:TotalPin} 
\end{equation} 
In the following we focus on WGMs with index $\mathbf{p}\equiv\{1,l_p,m, \text{TE}\}$ that are resonant with $\omega_{\mathrm{in}}$. This is allowed when the broadening of other WGMs is much smaller than their detuning from the input,\ i.e. $\left\vert \omega_{\mathbf{w}} - \omega_{\mathrm{in}}\right\vert \gg\bar{\kappa}_{\mathbf{w}}$ for $\omega_{\mathbf{w}}\neq\omega_{\mathrm{in}}$. While all WGMs with $m\leq l_p$ can be excited, WGMs with $\left\vert m\right\vert \approx l_p$ strongly dominate when the coupling is nearly adiabatic. 

Keeping the notation $\mathbf{q}\equiv\{\nu^{\prime},l^{\prime},m^{\prime}, \mathrm{TM}\}$, and $\boldsymbol{\alpha}\equiv\{\nu_s,l_s,m_s\}$ we now turn to the transmission power spectrum. 

\subsection{Transmission} \label{Sec:Tra}

Because of their relatively low frequencies, magnons typically have much smaller linewidths than the photons \cite{ZhangWGM16}, i.e. $\bar{\kappa}_{\boldsymbol{\mathbf{\alpha}}}\ll\kappa_{\mathbf{q}} + \bar{\kappa}_{\boldsymbol{\mathbf{q}}}$. In this limit $\chi_{\mathbf{q}}(\omega^{\prime}) \approx \chi_{\mathbf{q}}(\omega\pm\omega_{\boldsymbol{\alpha}})$ in $S_{\boldsymbol{\alpha}}^{\pm}$ in Eq. (\ref{PT:Gen}) such that 
\begin{equation}
	 \frac{P_{\mathrm{T}}^{\mathrm{TM}}(\omega_{\mathrm{out}})}{\bar{P}_{\mathrm{in}}} = \sum_{\boldsymbol{\alpha}\in\mathrm{SL}}\left[ \frac{\bar{\kappa}_{\boldsymbol{\mathbf{\alpha}}}S_{\boldsymbol{\alpha}}^+n_{\boldsymbol{\alpha}}}{\Delta_+^2 + \bar{\kappa}_{\boldsymbol{\mathbf{\alpha}}}^2/4} + \frac{\bar{\kappa}_{\boldsymbol{\mathbf{\alpha}}}S_{\boldsymbol{\alpha}}^- (n_{\boldsymbol{\alpha}} + 1)}{\Delta_-^2 + \bar{\kappa}_{\boldsymbol{\mathbf{\alpha}}}^2/4}\right] ,\label{PT:TEinp} 
\end{equation} 
where $\omega_{\mathrm{out}}$ is the center frequency of the detector (assumed to contain a filter of a small width) and, 
\begin{equation}
	 S_{\boldsymbol{\alpha}}^{\pm} = \left\vert \sum_{\mathbf{p},\mathbf{q}} \frac{\sqrt{\kappa_{\mathbf{p}}\kappa_{\mathbf{q}}}}{\bar{\kappa}_{\boldsymbol{\mathbf{p}}} + \kappa_{\mathbf{p}}}\frac{G_{\boldsymbol{\mathbf{pq\alpha}}}^{\pm}}{\delta_{\boldsymbol{\mathbf{q\alpha}}}^{\pm} - i\left( \bar{\kappa}_{\boldsymbol{\mathbf{q}}} + \kappa_{\mathbf{q}}\right) /2}\right\vert^2 . \label{Spm:TEinp} 
\end{equation} 
Here the sum over $\mathbf{p}$ refers to the sum over $m$ in the family of WGMs with frequency $\omega_{\mathbf{p}} = \omega_{\mathrm{in}}$, where the latter has been defined in Eq. (\ref{Def:Pin}), while $\delta_{\boldsymbol{\mathbf{q\alpha}}}^{\pm} = \omega_{\mathbf{q}} - (\omega_{\mathrm{in}}\pm\omega_{\boldsymbol{\alpha}})$ and $\Delta_{\pm} = \omega_{\mathrm{out}} - (\omega_{\mathrm{in}}\pm\omega_{\boldsymbol{\alpha}})$ are the frequency detunings of the output WGM and the output photon in the detector from the resonance, respectively. The scattering is efficient if both are less than the typical linewidths of WGMs. $P_{\mathrm{T}}^{\mathrm{TE}} = 0$ since $\text{TE} \rightarrow\text{TE}$ scattering is forbidden. 

\begin{figure}[ptb]
\begin{equation*}
\includegraphics[width=.47
\textwidth,keepaspectratio]{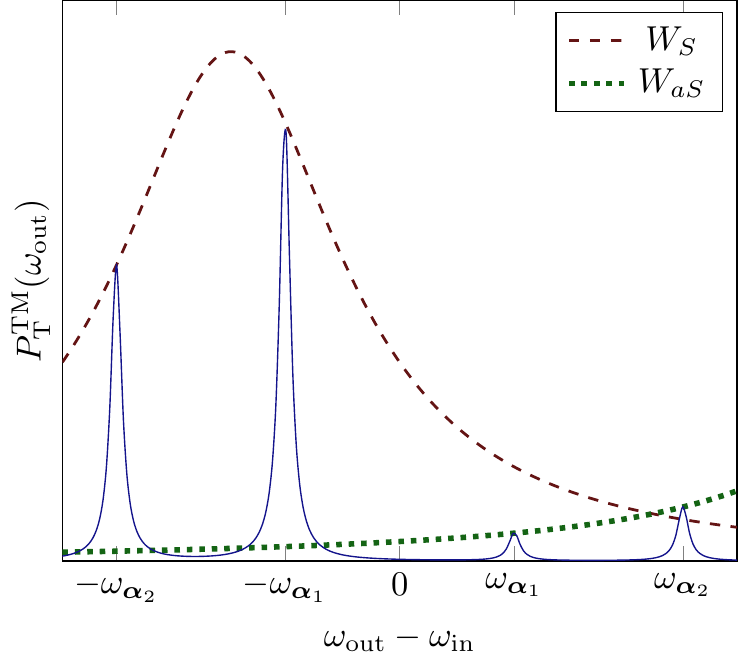}
\end{equation*}
\caption{Schematic inelastic scattering lineshape with two contributing magnon
$\boldsymbol{\alpha}_{1}$ and $\boldsymbol{\alpha}_{2} $ with constant
coupling $G^{\pm}$. The spectra can be written as the convolution of magnetic
and optical Lorentzian functions (solid line). The sharp peaks located
symmetrically around the origin are centered at the magnon mode frequencies
with broadening governed by the magnetization damping. The lines, $W_{S}$ and
$W_{aS}$ are envelope functions defined in the text. They are separated by
$2\Delta\omega_{\mathrm{\ FSR}}$ and broadened by optical decay rates. The
widths of all Lorentzians have been exaggerated for clarity. The actual
numbers are $\kappa_{\mathbf{S}, \mathbf{aS}}\sim2\pi\times1\,$GHz and
$\kappa_{\boldsymbol{ \alpha}}\sim2\pi\times1\,$MHz \cite{ZhangWGM16} implying
very small magnon lines inside a wide WGM Lorentzian.}
\label{Fig:Spectra}
\end{figure}

$S_{\boldsymbol{\alpha}}^{\pm}$ does not depend on $\omega_{\mathrm{out}}$ anymore, so each term in the sum of Eq. (\ref{PT:TEinp}) is a Lorentzian centered at $\omega_{\mathrm{out}} = \omega_{\mathrm{in}}\pm\omega_{\boldsymbol{\alpha}}$ (see Sec. \ref{Sec:Magnons}) with width $\bar{\kappa}_{\boldsymbol{\mathbf{\alpha}}}$ \cite{UsamiUnpub}. Each peak is well resolved if $\bar{\kappa}_{\boldsymbol{\mathbf{\alpha}}}<\left\vert \omega_{\boldsymbol{\alpha}} - \omega_{\boldsymbol{\alpha}^{\prime} \neq\boldsymbol{\alpha}}\right\vert $ [see Fig. \ref{Fig:Spectra}]. For small-$L$ magnons with $\left\vert \omega_{\boldsymbol{\alpha}} - \omega_{\boldsymbol{\alpha}^{\prime}}\right\vert \sim\gamma\mu_0M_s$ \cite{Fletcher59} this is the case when $\alpha_G\ll\mu_0M_s/B_0\sim0 . 1 - 1,$ which is easily fulfilled for YIG. We note that in previous experiments \cite{ZhangWGM16,Osada16,HaighWGM} the Kittel mode is selectively populated via microwave excitations ($n_{\mathbf{u}}\gg n_{\boldsymbol{\alpha \neq}\mathbf{u}}$) which overwhelms any other magnons, and thus only one peak was observed. 

The peak height at $\Delta_{\pm} = 0$ and integrated power $\bar{P}_{\mathrm{T}}$ are governed by the magnon linewidth ($\bar{\kappa}_{\boldsymbol{\mathbf{\alpha}}}$), magnon occupation ($n_{\boldsymbol{\alpha}}$), and $S_{\boldsymbol{\alpha}}^{\pm}$ (interpreted below). We may write 
\begin{equation}
	 \bar{P}_{\mathrm{T}} = \int\frac{d\omega^{\prime}}{2\pi}P_{\mathrm{T}}^{\mathrm{TM}}(\omega^{\prime}) = \sum_{\boldsymbol{\alpha}\in\mathrm{SL}}\left[ \bar{P}_{\boldsymbol{\alpha}}^+ + \bar{P}_{\boldsymbol{\alpha}}^-\right] ,\label{PT:Dec} 
\end{equation} 
where $\bar{P}_{\boldsymbol{\alpha}}^- = S_{\boldsymbol{\alpha}}^-(n_{\boldsymbol{\alpha}} + 1)\bar{P}_{\mathrm{in}}$ and $\bar{P}_{\boldsymbol{\alpha}}^+ = S_{\boldsymbol{\alpha}}^+ n_{\boldsymbol{\alpha}}\bar{P}_{\mathrm{in}}$ is carried by photons that underwent Stokes and anti-Stokes scattering respectively by $\boldsymbol{\alpha}$-magnons, corresponding to the integral of $P_{\mathrm{T}}$ across individual peaks in Eq. (\ref{PT:TEinp}). We can therefore interpret $S_{\boldsymbol{\alpha}}^{\pm}$ as the photon scattering probability from the contribution of many processes $\mathbf{p}\rightarrow\mathbf{q}$ by the magnon mode $\boldsymbol{\alpha}$. In the following, we discuss $S$ first for the Kittel mode and then for other small-$L$ magnons. 

\begin{figure}[ptb]
\begin{equation*}
\includegraphics[width=.48\textwidth,keepaspectratio]{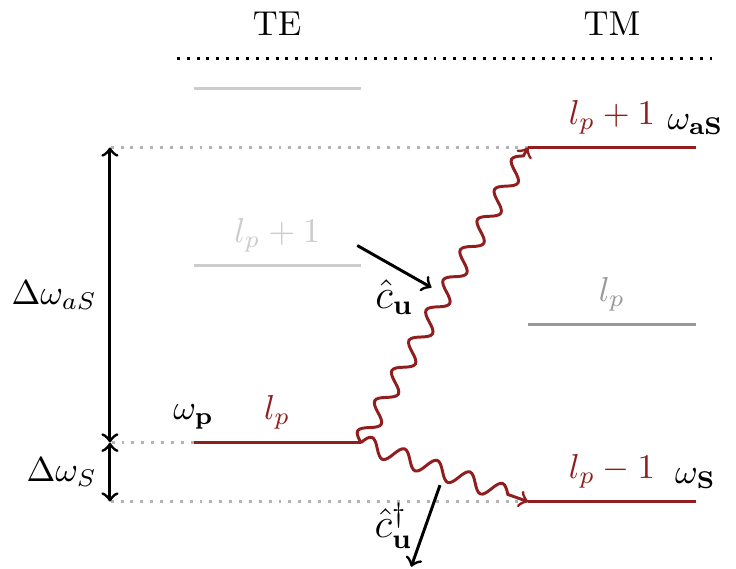}
\end{equation*}
\caption{Scattering by the Kittel magnon: Angular momentum conservation fixes
the final states [see Eq. (\ref{Coup:Kitt})]. The wavy lines denotes the
transitions associated with creation or annihilation of the Kittel mode.
Referring to Fig.\ref{WGMEnergies}, $\Delta\omega_{aS}=\Delta\omega
_{\mathrm{FSR}}-\Delta\omega_{ \mathrm{BF}}$ and $\Delta\omega_{S}
=\Delta\omega_{\mathrm{FSR }}+\Delta\omega_{\mathrm{BF}}$. Typical numbers are
$\Delta\omega_{S}\sim100-500$ GHz and $\Delta\omega_{aS}\sim1-20$ GHz for a
YIG radius of $a\sim100-500\,\mathrm{\mu m}$.}
\label{ScatKitt}
\end{figure}

\textit{Kittel mode:} The Kittel mode $\mathbf{u} = \{0,1,1\}$ can scatter a WGM $\mathbf{p} = \{1,l_p,m,\mathrm{TE}\}$ into either $\mathbf{S} = \{1,l_p - 1,m - 1,\mathrm{TM}\}$ or $\mathbf{aS} = \{1,l_p + 1,m + 1,\mathrm{TM}\} $ [see the selection rules in Eq. (\ref{Coup:Kitt})]. The optical transitions, valid for all $m$, are shown in Fig. \ref{ScatKitt}. In our example, $\Delta\omega_S = \Delta\omega_{\mathrm{FSR}} - \Delta \omega_{\mathrm{BF}} = 2\pi\times7 . 5$ GHz and $\Delta\omega_{aS} = \Delta\omega_{\mathrm{FSR}} + \Delta\omega_{\mathrm{BF}} = 2\pi\times209$ GHz. For magnon frequencies $\sim1 - 10$ GHz, the anti-Stokes scattering is highly non-resonant. 

By the magnetic field we can tune to the resonance condition $\omega_{\mathbf{u}} = \omega_{\mathbf{p}} - \omega_{\mathbf{S}} = \Delta \omega_{\mathrm{FSR}} - \Delta\omega_{\mathrm{BF}},$ where $\omega_{\mathbf{u}} = \gamma(B_0 - \mu_0M_s)$. Then, the Stokes scattering probability is maximized 
\begin{equation}
	 S_{\mathbf{u}}^- = |g_-|^2\left\vert \sum_m\frac{\sqrt{2\kappa_{\mathbf{p}}}}{\kappa_{\mathbf{p}} + \bar{\kappa}_{\boldsymbol{\mathbf{p}}}} \frac{\sqrt{2\kappa_{\mathbf{S}}}}{\kappa_{\mathbf{S}} + \bar{\kappa}_{\boldsymbol{\mathbf{S}}}}\right\vert^2 . 
\end{equation} 
The pre-factor $|g_-|^2$ is governed by the optomagnonic coupling in the sphere, while the second factor is a sum over the optical impedance matching parameters \cite{Biagioni_Antennas,Afzelius10} that determine the efficiency of the optical coupling. We find a lower bound for $S_{\mathbf{u}}^-$ by assuming that only the $m = l$ mode contributes. For $\bar{\kappa}_{\boldsymbol{\mathbf{p}}} = \bar{\kappa}_{\boldsymbol{\mathbf{S}}} = \kappa_{\mathbf{p}} = \kappa_{\mathbf{S}}$ with $m = l$ and an optical quality factor of $\omega_{\mathbf{p}}/\bar{\kappa}_{\boldsymbol{\mathbf{p}}} = 10^6$ comparable to experiments \cite{ZhangWGM16,HaighWGM}, $S_{\mathbf{u}}^- = 2\times10^{- 17}$. At $T = 300\,\mathrm{K}$, the number of magnons at $\omega_{\mathbf{u}} = 2\pi\times7$ GHz is $n_{\mathbf{u}} = 835,$ which leads to the scattered power of $\bar{P}_{\mathbf{u}}^-/\bar{P}_{\mathrm{in}} = 1 . 5\times10^{- 14}$. We note that the actual output power might be larger when more WGMs contribute to the above sum. We did not attempt to compute the power from a microscopic model of the optical coupling. 

For the same magnetic field, the anti-Stokes scattering is detuned from a resonance by $\omega_{\mathbf{aS}} - \omega_{\mathbf{p}} - \omega_{\mathbf{u}} = 2\Delta\omega_{\mathrm{BF}}$. For $2\Delta\omega_{\mathrm{BF}}\gg\bar{\kappa}_{\boldsymbol{\mathbf{aS}}},\kappa_{\mathbf{aS}}$, we obtain the S-aS intensity ratio 
\begin{equation}
	 \frac{\bar{P}_{\mathbf{u}}^-}{\bar{P}_{\mathbf{u}}^+} \approx \frac{n_{\mathbf{u}} + 1}{n_{\mathbf{u}}}\left\vert \frac{g_-}{g_+} \frac{2\Delta\omega_{\mathrm{BF}}}{\bar{\kappa}_{\boldsymbol{\mathbf{S}}} + \kappa_{\mathbf{S}}}\right\vert^2 . \label{Eq:SaS} 
\end{equation} 
Three mechanisms contribute to this ratio. The fraction $(n_{\mathbf{u}} + 1)/{n_{\mathbf{u}}}$ can be an important factor when $n_{\mathbf{u}}\lesssim1$ at low temperatures, but not at room temperature. The ratio of the microscopic scattering amplitudes 
\begin{equation}
	 \frac{g_-}{g_+} = \frac{\Psi_{LB}^{(2)} - \Psi_{CB}}{\Psi_{LB}^{(2)} + \Psi_{CB}} . 
\end{equation} 
can for instance be determined by BLS spectroscopy. Values in the range $0 . 1<\left\vert g_-/g_+\right\vert^2<10$ have been reported for YIG, depending on the magnetization direction and frequency \cite{Sandercock73,Wettling75}. For the parameters and configuration here, we find $g_-/g_+ = - 0 . 4 . $ The main reason for the observed large asymmetry \cite{Osada16,ZhangWGM16,HaighWGM} is therefore the non-resonant nature of the anti-Stokes scattering caused by the geometric and magnetic birefringence [see Eq. (\ref{LinBireF:Rep})]. Inserting the parameters introduced above, we find for the S-aS ratio $\bar{P}_{\mathbf{u}}^-/\bar{P}_{\mathbf{u}}^+ \approx S_{\mathbf{u}}^-/S_{\mathbf{u}}^+ = 2\times10^4 $. 

\textit{Small-L magnons:} We now discuss $S_{\boldsymbol{\alpha}}^{\pm}$, with $\boldsymbol{\alpha} = \{\nu_s,l_s,m_s\}$, for general small-$L$ magnons that gives the total power in each peak [see below Eq. (\ref{PT:Dec})]. Here we refrain from accurately computing the contributions from different magnon modes to the spectrum that have recently been observed \cite{UsamiUnpub}. Instead, we resort to making some qualitative observations. From Eq. (\ref{Spm:TEinp}), we see that $S_{\boldsymbol{\alpha}}$ involves a sum over all (symmetry-allowed) transition amplitudes, $\mathbf{p}\rightarrow \mathbf{q,}$ that in principle can cause interference effects. However, by choosing an appropriate magnetic field, the Stokes scattering is dominated strongly by the transition $l_p$ to $l_p - 1$ while other transitions are non-resonant. This can be done if the magnon band, $ = \gamma\mu_0M_s/2$ \cite{Walker57,Fletcher59}($\sim2\pi\times1$GHz for YIG), is not too large compared to the WGM linewidths (discussed in detail below). If this holds, we can ignore the non-resonant terms in the summation of Eq. (\ref{Spm:TEinp}) for $S_{\boldsymbol{\alpha}}^-$. Further, $G_{\boldsymbol{\mathbf{pq\alpha}}}^-$ is non-zero only if the $z$-component of angular momentum is conserved, i.e. $m^{\prime} = m - m_s$. Thus, for a given $\boldsymbol{\alpha}$ and $\mathbf{p} = \{1,l_p,m,\mathrm{TE}\}$ the WGM $\mathbf{S} = \{1,l_p - 1,m - m_s,\mathrm{TM}\}$ dominates. If $\kappa_{\mathbf{S}}\neq0$, we can observe this scattering. The anti-Stokes lines are caused by (non-resonant) scattering into different $\mathbf{aS} = \{1,l_{aS},m + m_s,\mathrm{TM}\}$ with $l_{aS}\geq l_p$ and $\left\vert m + m_s\right\vert \leq l_{aS}$ that can be calculated by Eq. (\ref{Spm:TEinp}). 

$S_{\boldsymbol{\alpha}}^{\pm}$ depends on the angular momenta and the energy of $\boldsymbol{\alpha}$ via the optomagnonic coupling $G$ and the detuning $\delta$ respectively [see Eq. (\ref{Spm:TEinp})]. BLS experiments are the method of choice to measure quasiparticle spectra, and this holds also for the present configuration. The WGMs transmission spectra sample the amplitude of the magnetization dynamics at the surfaces and are restricted by angular momentum conservation rules via the CG coefficients [ see Eqs. (\ref{Eq:GmTra}) and (\ref{Eq:GpTra})]. This implies that the total angular momentum should be approximately conserved, i.e. $G_{\boldsymbol{\mathbf{pS\alpha}}}^- \approx 0$ for $l_s\gg1$, restricting the number of optically active magnons; for $l_s\geq10$ peak heights are estimated from the CG coefficients to be less than $2\%$ of the Kittel mode. The lowest lying peaks are expected for $\boldsymbol{\alpha =}\left\{0,2,2\right\} ,\left\{0,2,1\right\} ,\left\{1,1,1\right\} $, with frequencies discussed briefly in Sec. \ref{Sec:Magnons} . WGM spectroscopy should be sensitive to surface effects that affect the magnonic boundary conditions (assumed to be free here). Similar restrictions hold for anti-Stokes scattering. 

In $S_{\boldsymbol{\alpha}}^-$ the frequency $\left\vert \delta_{\boldsymbol{\mathbf{q\alpha}}}^-\right\vert = \left\vert \omega_{\mathrm{in}} - \omega_{\boldsymbol{\alpha}} - \omega_{\mathbf{S}}\right\vert $ is the degree of non-resonance. When all symmetry-allowed $G^-$ are the same, each magnon peak in the Stokes spectrum are proportional to the density of state of the $\mathbf{S}$-WGM at the peak center 
\begin{equation}
	 W_S\left( \omega_{\mathrm{out}}\right) \propto\frac{\kappa_{\mathbf{S}} + \bar{\kappa}_{\boldsymbol{\mathbf{S}}}}{4\left( \omega_{\mathrm{in}} - \Delta\omega_S - \omega_{\mathrm{out}}\right)^2 + (\kappa_{\mathbf{S}} + \bar{\kappa}_{\mathbf{S}})^2}, 
\end{equation} 
as shown in Fig. \ref{Fig:Spectra} (red dashed lines), where $\omega_{\mathrm{out}} = \omega_{\mathrm{in}} - \omega_{\boldsymbol{\alpha}}$. Here, $\Delta\omega_S$ is defined in Fig. \ref{ScatKitt}. Only magnons with frequencies in a window of the order $\pm\left( \kappa_{\mathbf{S}} + \bar{\kappa}_{\mathbf{S}}\right) $ around $\Delta\omega_S$ are observable. This shows that we can optimize the scattering by shifting the magnon frequency, via an applied field. 

The anti-Stokes scattering for $l_{aS} = l_p + 1$ is plotted schematically in Fig. \ref{Fig:Spectra}. Here the peak heights are proportional to 
\begin{equation}
	 W_{aS}\left( \omega_{\mathrm{out}}\right) \propto\frac{\kappa_{\mathbf{aS}} + \bar{\kappa}_{\boldsymbol{\mathbf{aS}}}}{4\left( \omega_{\mathrm{in}} + \Delta\omega_{aS} - \omega_{\mathrm{out}}\right)^2 + (\kappa_{\mathbf{aS}} + \bar{\kappa}_{\mathbf{aS}})^2}, 
\end{equation} 
is shown in Fig. \ref{Fig:Spectra} (green dotted lines) with $\omega_{\mathrm{out}} = \omega_{\mathrm{in}} + \omega_{\boldsymbol{\alpha}}$. Here $\Delta\omega_{aS}$ is defined in Fig. \ref{ScatKitt}. Similar formulas hold for other $l_{aS}$ adding up to the total anti-Stokes peaks. The total number of observable peaks depends on $G^+$ that will distort the Lorentzian envelope for large detunings. 

\subsection{Reflection} \label{Sec:Ref}

We now turn to the inelastically reflected power. In a sphere, the DE magnons are degenerate at $\omega_{\mathrm{DE}} = \gamma\left( B_0 - 5 \mu_0 M_s/6\right) $ \cite{Walker57,DamEshSurface}. Therefore, only one Stokes peak is expected, to which the scattering amplitudes of all DE magnons contribute. Eq. (\ref{PT:Gen}) can then be simplified to 
\begin{equation}
	 \frac{P_{\mathrm{R}}^{\text{\textrm{TM}}}(\omega_{\mathrm{out}})}{\bar{P}_{\mathrm{in}}} = \frac{\bar{\kappa}_{\mathrm{DE}}(n_{\mathrm{DE}} + 1)}{\left( \omega_{\mathrm{out}} - \omega_{\text{\textrm{in}}} + \omega_{\mathrm{DE}}\right)^2 + \bar{\kappa}_{\mathrm{DE}}^2/4} \sum_{\boldsymbol{\alpha}\in\mathrm{LL}}S_{\boldsymbol{\alpha}}^- . \label{PR:TEinp} 
\end{equation} 
where $\boldsymbol{\alpha =}\left\{\nu_s,l_s\gg1,l_s\geq m_s \gg1\right\} ,$ $\omega_{\boldsymbol{\alpha}} = \omega_{\mathrm{DE}}$, $\kappa_{\boldsymbol{\alpha}} = \kappa_{\mathrm{DE}} = \alpha_G \omega_{\mathrm{DE}}$ and, at elevated temperatures, $n_{\boldsymbol{\alpha}} = n_{\mathrm{DE}} = k_BT/(\hbar\omega_{\boldsymbol{\alpha}})$. $P_{\mathrm{R}}^{\text{\textrm{TE}}} = 0$ since $\text{TE}\rightarrow \text{TE}$ scattering is forbidden. $P_{\mathrm{R}}$ is a Lorentzian centered at $\omega_{\mathbf{p}} - \omega_{\mathrm{DE}}$ with a width $\kappa_{\mathrm{DE}}$. The total integrated power over the peak 
\begin{equation}
	 \bar{P}_{\mathrm{R}}\overset{\triangle}{=}\int\frac{d\omega_{\mathrm{out}}}{2\pi}P_{\mathrm{R}}^{\text{\textrm{TM}}}(\omega^{\prime}), 
\end{equation} 
is then 
\begin{equation}
	 \frac{\bar{P}_{\mathrm{R}}}{\bar{P}_{\text{\textrm{in}}}} = \sum_{\boldsymbol{\alpha}\in\mathrm{LL}}\left\vert \sum_{\mathbf{p},\mathbf{q}}\frac{2\sqrt{\kappa_{\mathbf{p}}\kappa_{\mathbf{q}}}}{\bar{\kappa}_{\boldsymbol{\mathbf{p}}} + \kappa_{\mathbf{p}}}\frac{G_{\boldsymbol{\mathbf{pq\alpha}}}^-\sqrt{n_{\mathrm{DE}} + 1}}{i\delta_{\boldsymbol{\mathbf{q\alpha}}}^- + (\bar{\kappa}_{\boldsymbol{\mathbf{q}}} + \kappa_{\mathbf{q}})/2}\right\vert^2 . \label{RefSum} 
\end{equation} 
The summation over $\mathbf{p}$ includes all allowed $m>0$, while the $\mathbf{q}$ modes circle in the opposite direction $m^{\prime}<0$. 

\begin{figure}[ptb]
\begin{equation*}
\includegraphics[width=.38\textwidth,keepaspectratio]{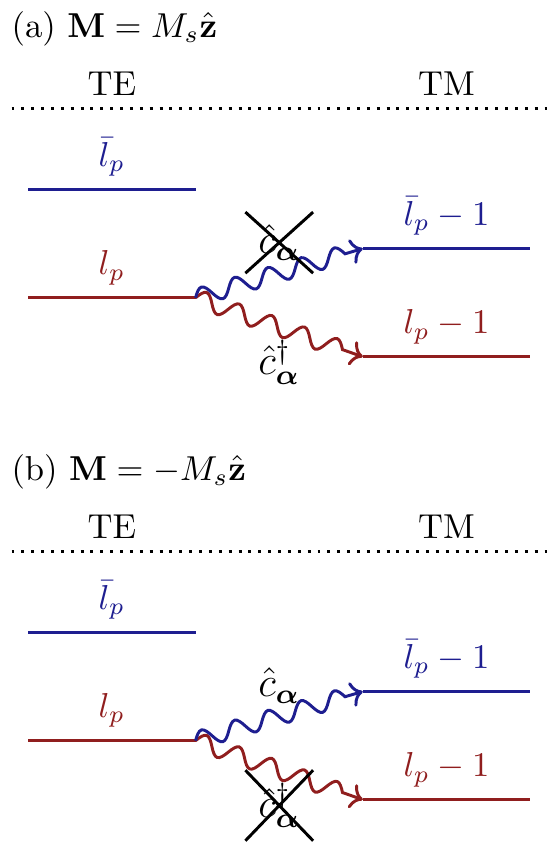}
\end{equation*}
\caption{Reflection by a Damon-Eshbach mode for (a) $\mathbf{M}=M_{s}\hat{
\boldsymbol{\mathbf{z}}}$ and (b) $\mathbf{M}-M_{s}\hat{\boldsymbol{\mathbf{z
}}}$. Due to the chirality of DE magnons, one of the Stokes or anti-Stokes
transitions is forbidden, depending on the direction of the magnetization.}
\label{ScatDE}
\end{figure}

Using Fig. \ref{WGMEnergies}, we see that $\mathbf{q} = \{1,l_p - 1,m^{\prime}, \mathrm{TM}\}$ for $m^{\prime} = m - m_s<0$ are the only resonant final states for magnon frequencies $<2\pi\times20\,$GHz [illustrated in Fig. \ref{ScatDE}(a)]. A magnetic field ($\sim1\,\mathrm{T}$ ) can tune the system into the resonant condition $\omega_{\mathrm{DE}} = \Delta \omega_{\mathrm{FSR}} - \Delta\omega_{\mathrm{BF}}$. We can estimate a lower bound of the output power by assuming that only $m = l_p$ and $m^{\prime} = - \left( l_p - 1\right) $ modes couple to the fiber and the magnons with $m_s = l_s = 2l_p - 1$ dominate. There is only one state $\boldsymbol{\alpha}$ with $m_s = l_s$ labeled as $\nu_s = 0$ [see Sec. \ref{Sec:Magnons}], hence 
\begin{equation}
	 \frac{\bar{P}_{\mathrm{R}}}{\bar{P}_{\mathrm{in}}} = \frac{2\kappa_{\mathbf{p}}}{\left( \kappa_{\mathbf{p}} + \bar{\kappa}_{\boldsymbol{\mathbf{p}}}\right)^2}\frac{2\kappa_{\mathbf{q}}}{\left( \kappa_{\mathbf{q}} + \bar{\kappa}_{\boldsymbol{\mathbf{q}}}\right)^2}\left\vert g_-\Xi_-\right\vert^2\left( n_{\mathrm{DE}} + 1\right) , 
\end{equation} 
where $\Xi_-$ has been defined in Eq. (\ref{Prefm}). Using the parameters for YIG\ given above, we arrive at the estimate $\bar{P}_{\mathrm{R}}/\bar{P}_{\mathrm{in}} = 4\times10^{- 13}$ at $T = 300$\,K. The actual output power will be larger, depending on the optical coupling and multiple contributing DE magnons. 

Since the WGMs are spatially extended compared to the DE modes, the radial overlap interface for $\nu_s\neq0$ is suppressed. While for small-$L$ magnons with approximately constant amplitude, the orthogonality of the WGMs efficiently suppresses inter-branch scattering with $\nu\neq\nu^{\prime},$ the DE modes are localized to the surface, which allows scattering between WGMs $\nu^{\prime}\neq\nu$. In the present configuration with $\mathbf{M} \approx M_s\hat{\boldsymbol{\mathbf{z}}}$ the $\nu = 1$ intra-branch scattering dominates, because $\nu^{\prime}\neq\nu$ transition are non-resonant (see Fig. \ref{ScatDE}). The situation is different for $\mathbf{M} \approx - M_s \hat{\boldsymbol{\mathbf{z}}}$ (see next section). 

The angular momentum of the WGMs for infrared light is\ typically of the order of $l_p\sim10^3 . $ The DE magnons that reflect these photons have angular moments of the same order. In YIG spheres with $a\sim100\,\mathrm{\mu m}$ exchange effects become significant only for $l_s>10^4$, implying that we can neglect exchange. For exchange energy smaller than the magnetostatic energy, $\sim\gamma\mu_0M_s$, we expect the magnons to be chiral still, but with a different magnon spatial distribution affecting the overlap. 

\begin{figure*}[ptb]
\begin{equation*}
\includegraphics[width=\textwidth,keepaspectratio]{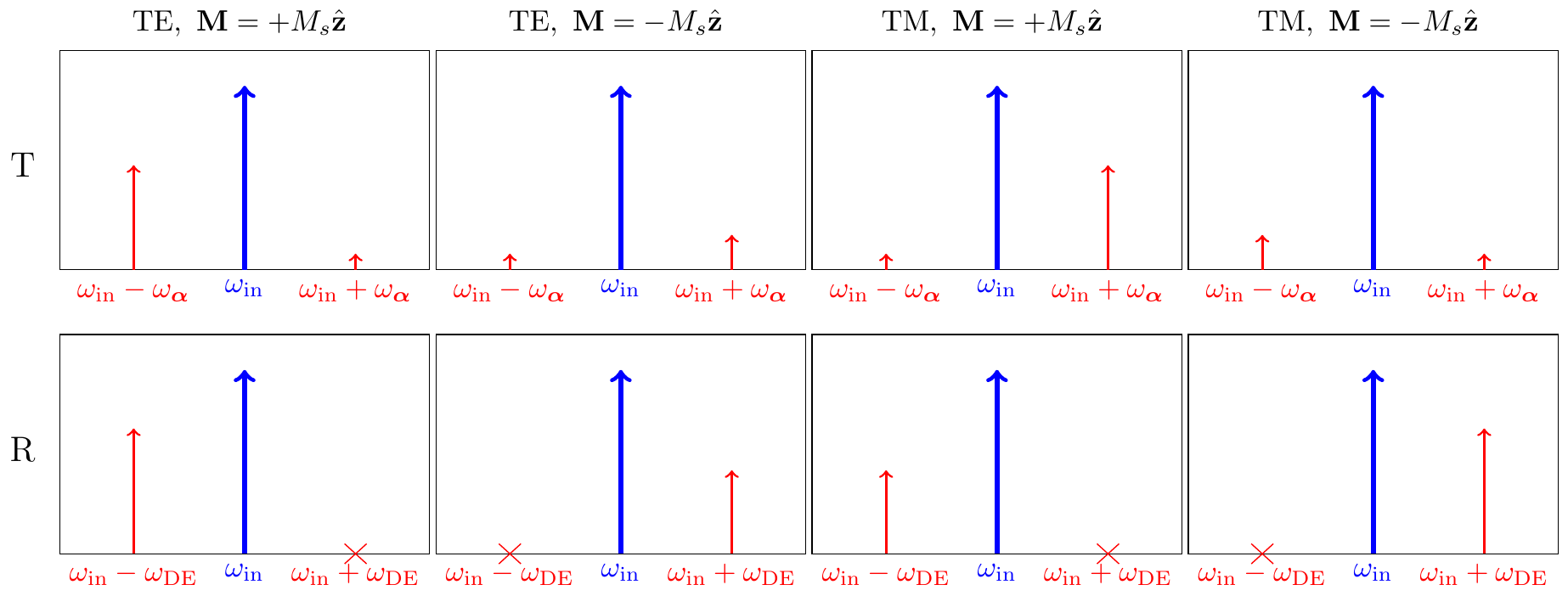}
\end{equation*}
\caption{Schematic light scattering spectra by a magnetic sphere in proximity
to a single mode optical fiber, emphasizing the S-aS asymmetry. The blue arrow
marks the frequency $\omega_{\text{in}}$ of light input and the panel T(R)
denotes the transmission(reflection) spectra with Stokes and anti-Stokes
lines. $\omega_{\boldsymbol{\alpha}}$ and $\omega_{\mathrm{DE}}$ are the
resonance frequencies of magnons involved in the transmission and reflection,
respectively. Each column corresponds to one of the four cases, $\sigma
\in\{\text{TE},\text{TM} \}$ or $\mathbf{M}\parallel\pm\hat
{\boldsymbol{\mathbf{z}}}$. A cross denotes complete absence of a peak by
chirality selection. The peak heights are not on scale, but peaks with the
same height have power in the same order-of-magnitude. We assume that the
input frequency and the magnetic field have been tuned to the resonant
scattering condition in each case as discussed in the main text.}
\label{Res}
\end{figure*}

\section{Other configurations} \label{Sec:Other}
 The above analysis focuses on a TE polarized incident photons in Fig. \ref{Setup} that couple to the $\nu = 1$ WGM. We now briefly discuss other configurations involving a different WGM, magnetization direction, and polarization. The conclusions are summarized in Fig. \ref{Res} (not to scale). 

\textit{Other WGMs:} Magnons close to the Kittel mode with small angular momentum and nearly constant amplitude over the sphere\ can scatter WGMs with the same number of nodes $\nu = \nu^{\prime}$ only. The $\nu = 1$ mode is expected to dominate because of the larger evanescent coupling. DE magnons may form an exception, since on magnetization reversal interbranch scattering should become observable (next paragraph). 

\textit{Magnetization:} The results for $\mathbf{M} \approx M_s \hat{\boldsymbol{\mathbf{z}}}$ can be used to understand the scattering after magnetization reversal $\mathbf{M} \approx - M_s\hat{\boldsymbol{\mathbf{z}}}$ . This inverts the magnon angular momenta $m_s$ in the selection rules, which might lead to a na\"{\i}ve expectation that we simply have to exchange Stokes and anti-Stokes scattering. This is not the case, however, as we discuss now for the Kittel mode. For simplicity, let's consider only the dominant optical mode with $m = l_p$ which has simple selection rules, viz. $l_p\rightarrow l_p\pm1$. The Stokes scattering, in this case, occurs from $l_p$ to $l_p + 1$ because of angular momentum conservation. This has a large detuning of $\Delta\omega_{aS} + \omega_{\boldsymbol{\alpha}}$ [$\Delta\omega_{aS}$ defined in Fig. \ref{ScatKitt}]. The anti-Stokes transition $l_p\rightarrow l_p - 1$ (consistent with angular momentum conservation) is also non-resonant with a detuning of $\Delta\omega_S + \omega_{\boldsymbol{\alpha}}$. Therefore, the reduction of angular momentum of WGM is accompanied by an increase in energy and vice versa, which is not favourable for scattering [see Fig \ref{WGMEnergies}], decreasing both the peaks. For the parameters of the YIG sphere used before, we find in this case $\bar{P}_{\boldsymbol{u}}^+/\bar{P}_{\mathrm{in}}\sim10^{- 16}$ and a weaker (inverted) S-aS asymmetry $\bar{P}_{\boldsymbol{u}}^+/\bar{P}_{\boldsymbol{u}}^-\sim100$. 

The chirality of DE magnons is reversed with magnetization, and anti-Stokes scattering becomes allowed while the Stokes scattering is forbidden. Using Fig. \ref{ScatDE}, we see that the scattering cannot be resonant now for intra-branch scattering. However, the WGM of the $\nu = 2$ branch is close in frequency and we can choose $\omega_{\mathrm{DE}} = \Delta\omega_{\mathrm{BF}} - \Delta\omega_{\mathrm{FSR}} + \Delta\omega_{\mathrm{RG}} = 2\pi\times8 . 8$ GHz in our case. Therefore, we can still have resonant reflection under certain conditions, i.e. the evanescent coupling of $\nu = 2$ branch is significant and the above detuning is not too large. 

\textit{TM input:} All the arguments, so far, can be repeated for the case of TM polarized light input. We can understand the schematic results in Fig. \ref{Res} by time-reversal symmetry arguments. The time-reversal operator, $\mathcal{T}$, inverts the magnetization and the WGM circulation direction, so the direction of a WGM w.r.t the magnetization is conserved. $\mathcal{T}$ exchanges magnon annihilation and creation as well as input WGM and scattered WGM. The last condition implies that both the polarization and the direction of motion of WGMs must be interchanged. Thus, the transmission spectrum for TM input when $\mathbf{M} = \pm M_s\hat{\boldsymbol{\mathbf{z}}} $ is recovered by interchanging Stokes and anti-Stokes scattering in the transmission of TE input when $\mathbf{M} = \pm M_s\hat{\boldsymbol{\mathbf{z}}}$. On the other hand, the reflection spectrum for $\{\text{TE},\ \mathbf{M} = \pm M_s \hat{\boldsymbol{\mathbf{z}}}\}$ is mirror symmetric (across the input frequency) with $\{\text{TM},\ \mathbf{M} = \mp M_s\hat{\boldsymbol{\mathbf{z}}}\}$. 

%%%%%% Conclusion 

\section{Discussion and outlook} \label{Sec:Conc}

Our theory demonstrates that the transmission (reflection) spectra of inelastically scattered photon involves magnons with small (large) angular momentum. Both show a pronounced asymmetry in the probability of Stokes and the anti-Stokes scattering, albeit for very different reasons (discussed below). The conclusions hold for (approximately) spherical cavities with magnetizations perpendicular to the WGM orbits. Here we briefly discuss non-spherical samples and general magnetization directions. In the following $\mathbf{L}_W$ and $\mathbf{L}_m$ denote the angular momenta of WGMs and magnons, respectively. 

Our finding that the pronounced S-aS asymmetry observed in forward scattering \cite{ZhangWGM16,Osada16,HaighWGM} is caused by linear birefringence agrees with that of \cite{HaighWGM}. However, it is at odds with \cite{Osada16} and \cite{ZhangWGM16}, who attribute the asymmetry to partial elliptical polarization of the evanescent tails of WGMs outside the sphere \cite{Junge13} (that we believe should interact only very weakly with the magnetization). The present results can be carried to dielectric shapes with closed extreme orbits with sufficiently weak curvature, such as bottlenecks, rings, etc. This implies that the forward scattering power is increased for oblate ellipsoid with smaller volume and equal curvature at the equator, increasing the relative overlap volume [see Eq. (\ref{Coup:Kitt})]. 

To the best of our knowledge, back scattering of light by magnons has not yet been discussed in the literature. We find a perfect selection rule for S-aS back-scattering by DE magnotastic spin waves. The physical reason is their chirality that locks the sign of the orbital angular momenta to the magnetization direction. The interactions between DE magnons and WGMs is enhanced because (i) they are confined to the same equatorial region of the sphere and (ii) the dwell time in which the interaction can take place is long when dissipation is weak. While the reflection power estimated here is not yet very high, engineering the spatial distribution of WGMs might lead to the coveted strong interaction between light and magnetism. We will discuss such optimizations in a future article. 

The analysis for arbitrary magnetization directions is more complex because the rotational symmetry about $z$-axis is broken. In the case of transmission $\left\vert \mathbf{L}_m\right\vert \ll\left\vert \mathbf{L}_W\right\vert $, so the WGMs are only slightly changed after scattering. Thus we expect the spatial overlap to be the same as our work. The CG coefficients have to be generalized for non-collinear magnetization changing the selection rules but will perhaps not change the order of magnitude of the coupling constants $G$. The change in selection rules will significantly affect the detunings, thus the peak heights and the S-aS asymmetry, which we believe to be the major change. As we discussed before, inverting the magnetization inverts the S-aS asymmetry. For the particular case of in-plane magnetization, say $\mathbf{M} = M_s\hat{\boldsymbol{\mathbf{x}}}$, the S-aS asymmetry in transmission is suppressed because the angular momentum $L_m\hat{\boldsymbol{\mathbf{x}}}$ of a spin-1/2 system is formed as a coherent linear superposition of $L_m\hat{\boldsymbol{\mathbf{z}}}$ and $ - L_m\hat{\boldsymbol{\mathbf{z}}}$ and adding or subtracting $L_m \hat{\boldsymbol{\mathbf{x}}}$ can generate resonant scattering from $l$ to $l - 1$ with equal probability. In the in-plane magnetization configuration the photons also experience inhomogeneous Faraday rotation and Cotton-Mouton ellipticity \cite{Lan11,Haigh15}. This causes a small ellipticity in the WGMs that contributes to the light scattering only to higher order. The treatment of light reflection for a general magnetization direction is fairly complicated as $\mathbf{L}_W$ changes significantly after scattering and therefore beyond the scope of this work. 

While (undoped) YIG has excellent magnetic quality, its magneto-optical effects are weak \cite{WettlingData76}. The best material for cavity optomagnonics would maximize $S^{\pm}$ [see Eq. (\ref{Spm:TEinp})]. It should have a window of high transparency, i.e. small optical losses (eventually by polishing the surface \cite{ZhangWGM16}), and large MO effects that bolster the $g_{\pm}$ [Eq. (\ref{Def:gpm})]. Provided that it is much smaller than the optical broadening, the magnon linewidth governed by the Gilbert damping does not play a role in the integrated scattered power. Doping YIG or substituting yttrium by magnetic rare earth atoms increases MO effects but may also lead to a deterioration of the optical and magnetic quality \cite{Mitchell87}. 

While we considered BLS by magnons, light can be scattered by other excitations, such as phonons. The latter generate inelastic scattering at $\omega_{\mathrm{in}}\pm\omega_{ph}$ where $\omega_{ph}$ are optically active phonon frequencies. Unless a phonon is resonant with any of the relevant magnons, the two scattering are independent and can be easily separated from the magnetic signals by e.g. changing the magnetic field. 

In summary, we studied BLS by magnons in spherical cavities, restricting to WGMs, with the magnetization perpendicular to WGM orbit. We expect our discussion of BLS by magnons in WGM cavities to hold for more general geometries, but not for a general magnetization direction. The expressions derived here can be used for improving the coupling between magnons and phonons. The dependence of the scattered power on the input mode as illustrated by Fig. \ref{Res}, allows controllable energy transfer between magnet and light. DE magnons can be pumped or annihilated selectively by reflection of light. Similarly, the low-$L$ magnons can be pumped or cooled by light depending on the polarization of the input. A forthcoming article is devoted to the thermodynamics of light scattering by magnetic spheres. 

We acknowledge helpful discussions with Koji Usami and Koji Satoh. This work is supported by the DFG Priority Programme 1538 \textquotedblleft Spin-Caloric Transport\textquotedblright, the NWO, and JSPS\ Grants-in-Aid for Scientific Research (Grant Nos. 25247056, 25220910, and 26103006). 

\appendix %dummy comment inserted by tex2lyx to ensure that this paragraph is not empty 

\section{Scattering amplitudes} \label{App:ScatAmp}

In this Appendix, we calculate the scattering amplitudes of a TE WGM into a TM WGM for both forward and back scattering. To this end we use the expressions for the optical and magnonic fields in Sec. \ref{Sec:Model} to calculate the integrals in Eqs. (\ref{Def:Gp}) and (\ref{Def:Gm}). 

\subsection{Normalization} 

We normalize the field amplitudes by equating the energy of an electromagnetic mode $\mathbf{p}\equiv\{\nu,l,m,\sigma\}$ to that of a harmonic oscillator with frequency $\omega_{\mathbf{p}} = ck_{\mathbf{p}}/n_{\sigma}$, 
\begin{equation}
	 \int\left( \frac{\epsilon_{\sigma}}{2}\left\vert \mathbf{E}_p\right\vert^2 + \frac{1}{2\mu_0}\left\vert \mathbf{B}_p\right\vert^2\right) d \mathbf{r} = \frac{\hbar\omega_{\mathbf{p}}}{2}, 
\end{equation} 
where $\mathbf{E}_{\mathbf{p}}$ is given by Eqs. (\ref{ETE}) and (\ref{ETM}) for either polarization, $i\omega_{\mathbf{p}}\mathbf{B}_{\mathbf{p}} = \boldsymbol{\nabla}\times\mathbf{E}_{\mathbf{p}}$, and $\epsilon_{\sigma} = \epsilon_0n_{\sigma}^2$. The contribution of fields outside the sphere to the energy is negligible owing to a smaller amplitude and fast radial decay, $\sim a/l$. Using the asymptotic form for large $l$ and $r \approx a$ with $k_{\mathbf{p}}a \approx l$ [see Eq. (\ref{ResWGM})] 
\begin{equation}
	 \frac{\hbar\omega_{\mathbf{p}}}{2} = \frac{\epsilon_{\sigma}\mathcal{E}_{\mathbf{p}}^2}{2}\int\left[ \left\vert j_l\right\vert^2\left( \left\vert \mathbf{Y}_l^m\right\vert^2 + \left\vert \mathbf{X}_l^m\right\vert^2\right) + \left\vert (j_l)^{\prime}\right\vert^2\left\vert \mathbf{Z}_l^m\right\vert^2\right] d\mathbf{r}, 
\end{equation} 
where $j_l\equiv j_l(k_{\mathbf{p}}r)$ is a spherical Bessel function, $(j_l)^{\prime}$ its first derivative, and the integral is only within $r<a$ . $\{\mathbf{X},\mathbf{Y},\mathbf{Z}\}$ are vector spherical harmonics (VSH) defined by 
\begin{align}
	 \mathbf{X}_l^m &= Y_l^m\hat{\mathbf{r}},\\
	 \mathbf{Y}_l^m &= \frac{1}{\sqrt{l(l + 1)}}\mathbf{L}Y_l^m,\\ \mathbf{Z}_l^m &= \hat{\mathbf{r}}\times\mathbf{Y}_l^m, 
\end{align} 
and $Y_l^m$ are scalar spherical harmonics (SH), 
\begin{equation}
	 Y_l^m(\theta,\phi) = ( - 1)^m \sqrt{\frac{2l + 1}{4\pi}\frac{(l - m)!}{(l + m)!}} P_l^m(\cos\theta)e^{im\phi} . \label{Exp:Ylm} 
\end{equation} 
in terms of the associated Legendre polynomials 
\begin{equation}
	 P_l^m(x) = \frac{( - 1)^m}{2^ll!}\left( 1 - x^2\right)^{m/2} \frac{d^{l + m}}{dx^{l + m}}\left( x^2 - 1\right)^l . 
\end{equation} 
After carrying out the angular integrals and using $\left\vert j_l^{\prime}\right\vert^2\ll|j_l|^2$ for large $l$ 
\begin{equation}
	 \frac{\hbar\omega_{\mathbf{p}}}{2} \approx \epsilon_{\sigma}\mathcal{E}_{\mathbf{p}}^2\int_0^a\left\vert j_l\left( k_{\mathbf{p}}r\right) \right\vert^2r^2dr, 
\end{equation}

We may further simplify results by $k_{\mathbf{p}}r = l + t(l/2)^{1/3}$, where $t\leq\beta_{\nu}$ as seen from Eq. (\ref{ResWGM}), and the asymptotic form of the Bessel function \cite{AbrSteg}, 
\begin{equation}
	 j_l\left( l + t\left( \frac{l}{2}\right)^{1/3}\right) \approx \sqrt{\frac{\pi}{2l}}\left( \frac{2}{l}\right)^{1/3}\mathrm{Ai}( - t),\label{jlAiry} 
\end{equation} 
where $\text{Ai}$ is the Airy function. Thus 
\begin{equation}
	 \frac{\hbar\omega_{\mathbf{p}}}{2} \approx \frac{\pi\epsilon_{\sigma}\mathcal{E}_{\mathbf{p}}^2}{k_{\mathbf{p}}^3}\left( \frac{l}{2}\right)^{2/3} \int_0^{\infty}\mathrm{Ai}^2(t - \beta_{\nu})dt . 
\end{equation} 
Inserting $\int\mathrm{Ai}^2(t)dt = t\mathrm{Ai}^2(t) - \mathrm{Ai}^{\prime2}(t)$, 
\begin{equation}
	 \mathcal{E}_{\mathbf{p}} \approx \frac{4}{a^2\left\vert \mathrm{Ai}^{\prime}( - \beta_{\nu})\right\vert}\sqrt{\frac{\hbar c}{2n_{\sigma}\pi\epsilon_{\sigma}}}\left( \frac{l}{2}\right)^{5/3}, 
\end{equation} 
which is the desired normalization factor for a single photon in a WGM. 

We normalize the magnetization to the spin angular momentum as $\gamma\hbar = \int d\mathbf{r}\left( M_s - M_{z,\boldsymbol{\alpha}}\right) $, or equivalently, 
\begin{equation}
	 \gamma\hbar \approx \int\frac{|M_{+ ,\boldsymbol{\alpha}}|^2 + |M_{- , \boldsymbol{\alpha}}|^2}{4M_s}d\mathbf{r} . 
\end{equation} 
For the Kittel mode, the magnetization is constant with $M_+(\mathbf{r}) = 0$ and $M_-(\mathbf{r}) = M_u$ giving 
\begin{equation}
	 M_u = \sqrt{\frac{4\gamma\hbar M_s}{V}} . \label{Val:Mu} 
\end{equation} 
Using Eq. (\ref{DE:Dist}) for the DE magnons, we get 
\begin{equation}
	 \frac{4\gamma\hbar M_s}{a^3} = 2\pi M_{l_s}^2\int_0^1\rho^{2l_s}d\rho\int_0^{\pi}\sin^{2l_s - 1} \theta\ d\theta, 
\end{equation} 
where $\rho = r/a$ is the normalized radial coordinate. Performing the integrals, 
\begin{equation}
	 M_{l_s} = \left( \frac{l_s}{\pi}\right)^{3/4}\sqrt{\frac{4\gamma\hbar M_s}{a^3}},\label{Val:Mls} 
\end{equation} 
normalizing the DE modes. 

\subsection{Kittel mode} \label{App:KittScat}

Here we evaluate the integrals Eqs. (\ref{Def:Gp}) and (\ref{Def:Gm}) for a WGM with index $\mathbf{p}\equiv\{\nu,l,m,TE\}$ that scatters into $\mathbf{q}\equiv\{r^{\prime},l^{\prime},m^{\prime},TM\}$ by the Kittel mode, $\boldsymbol{\alpha}\equiv\{0,1,1\}$. Throughout this section $l,m\gg1,|l - m|$ and $l^{\prime},m^{\prime}\gg1,|l^{\prime} - m^{\prime}|$. The coupling integrals, Eqs. (\ref{Def:Gp}) and (\ref{Def:Gm}), can be written as 
\begin{equation}
	 G_{\boldsymbol{\mathbf{pq\alpha}}}^{\pm} = \frac{\mathcal{G}_{\pm}M_u}{4\hbar}I_{\pm}, 
\end{equation} 
where 
\begin{equation}
	 I_{\pm} = \int_{|r|<a} E_{\mathbf{p},z}E_{\mathbf{q},\pm}^{\ast}d\mathbf{r} . 
\end{equation} 
Putting the distribution of electric fields, Eqs. (\ref{ETE}) and (\ref{ETM}), we get 
\begin{equation}
	 I_{\pm} \approx i\mathcal{E}_{\mathbf{p}}\mathcal{E}_{\mathbf{q}}\int\left( j_l(k_{\mathbf{p}}r)Y_l^m\right) \left( j_{l^{\prime}}(k_{\mathbf{q}}r)\sin\theta e^{\pm i\phi}(Y_{l^{\prime}}^{m^{\prime}})^{\ast}\right) d\mathbf{r} . 
\end{equation} 
For large $l$, we have a recursive relation, $\sin\theta e^{\pm i\phi}Y_{l^{\prime}}^{m^{\prime}} = Y_{l^{\prime}\pm1}^{m^{\prime}\pm1}$, giving 
\begin{align}
	 I_{\pm} & \approx i\mathcal{E}_{\mathbf{p}}\mathcal{E}_{\mathbf{q}}\int j_l(k_{\mathbf{p}}r)j_{l^{\prime}}(k_{\mathbf{q}}r)Y_l^m\left( Y_{l^{\prime}\mp1}^{m^{\prime}\mp1}\right)^{\ast}d\mathbf{r} \nonumber \\
	 &= i\mathcal{E}_{\mathbf{p}}\mathcal{E}_{\mathbf{q}}\delta_{l,l^{\prime} \mp1}\delta_{m,m^{\prime}\mp1}\int r^2drj_l(k_{\mathbf{p}}r)j_{l^{\prime}}(k_{\mathbf{q}}r) . \label{Ipm:AngDone} 
\end{align} 
where we used the orthonormality of the SHs. Since the integral is dominated by $r \approx a$, we can use the asymptotic form of the Bessel function, Eq. (\ref{jlAiry}) and the orthogonality relation 
\begin{equation}
	 \int_0^{\infty}dt\ \mathrm{Ai}(t - \beta_{\nu})\mathrm{Ai} (t - \beta_{\nu^{\prime}}) = \delta_{\nu,\nu^{\prime}}\left( \mathrm{Ai}^{\prime}( - \beta_{\nu})\right)^2, 
\end{equation} 
to arrive at 
\begin{equation}
	 I_{\pm} \approx i\frac{\hbar\omega_{\mathbf{p}}}{2\epsilon_s} \delta_{\nu,\nu^{\prime}}\delta_{l,l^{\prime}\mp1}\delta_{m,m^{\prime}\mp1}, 
\end{equation} 
for $\epsilon_s\gg\{fM_s,gM_s^2,g^{\prime}M_s^2\}$. Finally, 
\begin{equation}
	 G_{\mathbf{pq}\boldsymbol{\alpha}}^{\pm} = \frac{gM_s\pm f}{2\epsilon_s} \frac{M_s}{\sqrt{sV}}i\omega_p\delta_{\nu,\nu^{\prime}}\delta_{l,l^{\prime}\mp1}\delta_{m,m^{\prime}\mp1}, 
\end{equation} 
which can be written in terms of the MO constants defined in Eqs. (\ref{CB}) - (\ref{LB2}) and lead to Eq. (\ref{Coup:Kitt}). The orthonormality of the SHs reflects the conservation of angular momentum and the orthonormality of WGMs in the radial direction leads to the radial selection rule. 

\subsection{DE modes} \label{App:DEScat}

We calculate scattering of a WGM with $\mathbf{p}\equiv\{\nu,l,m,TE\}$ into one with index $\mathbf{q}\equiv\{\nu^{\prime},l^{\prime},m^{\prime},TM\}$ by a particular DE magnon given by $\boldsymbol{\alpha}\equiv\{0,l_s,l_s \}$ . We take the case of $m>0$ which implies $m^{\prime}<0$ as discussed in the main text, Sec. \ref{Coup:LL}. Here, we assume $l,m\gg1,\left\vert l - m\right\vert $, and similarly $l^{\prime},|m^{\prime}|\gg1,|l^{\prime} + m^{\prime}|$. The coupling constants, Eqs. (\ref{Def:Gp}) and (\ref{Def:Gm}), are 
\begin{equation}
	 \hbar G_{\mathbf{pq}\boldsymbol{\alpha}}^{\pm} = \frac{i\mathcal{G}_{\pm} \mathcal{E}_{\mathbf{p}}\mathcal{E}_{\mathbf{q}}M_{l_s}a^3}{4}\mathcal{R} \Theta_{\pm},\label{Eq:GpmInt} 
\end{equation} 
where we divided the integrals into the angular ($\Theta$) and the radial ($\mathcal{R}$) parts. The angular integral is 
\begin{equation}
	 \Theta_{\pm} = \int d\Omega Y_l^m\left[ \sin\theta e^{\pm i\phi}\left( Y_{l^{\prime}}^{m^{\prime}}\right)^{\ast}\right] \left( \sin\theta e^{\pm i\phi}\right)^{l_s - 1},\label{Int:AngDE} 
\end{equation} 
where $d\Omega = \sin\theta d\theta d\phi$ is the angular differential. As $Y_l^m\sim e^{im\phi}$, we get that $\Theta_{\pm}$ is non-zero only if $m - m^{\prime}\pm l_s = 0$. As discussed in the text, $m^{\prime} \approx - m$ , so $\Theta_+ = 0$. $\Theta_-$ can be evaluated by using $(Y_L^M)^{\ast} = ( - 1)^MY_L^{- M}$, 
\begin{equation}
	 Y_L^L \approx \frac{L^{1/4}}{\sqrt{2}\ \pi^{3/4}}\sin^L\theta e^{iL\phi}, 
\end{equation} 
and the identity, 
\begin{multline*}
\int d\Omega\ Y_{L_{1}}^{M_{1}}Y_{L_{2}}^{M_{2}}\left( Y_{L}^{M}\right)
^{\ast}\approx\sqrt{ \frac{L_{1}L_{2}}{2\pi L}}\\
\left\langle L_{1},M_{1};L_{2},M_{2}\right\vert \left. L_{3},M_{3}
\right\rangle \left\langle L_{1},0;L_{2},0\right\vert \left. L_{3}
,0\right\rangle ,
\end{multline*}
where the approximations holds for $L,L_i\gg1$ for $i\in\{1,2,3\}$. We get 
\begin{equation}
	 \Theta_- = \frac{\pi^{3/4}}{l_s^{3/4}}\sqrt{\frac{ll^{\prime}}{\pi}} \left\langle l,m;l^{\prime},|m^{\prime}|\right\vert \left . l_s ,m_s\right\rangle \left\langle l,0;l^{\prime},0\right\vert \left . l_s,0\right\rangle . 
\end{equation} 
The radial integral is 
\begin{equation}
	 \mathcal{R} = \int_0^1j_l(k_{\mathbf{p}}a\rho)\left[ j_{l^{\prime}}(k_{\mathbf{q}}a\rho) - j_{l^{\prime}}^{\prime}(k_{\mathbf{q}}a\rho)\right] \rho^{l_s + 1}d\rho,\label{Int:RadDE} 
\end{equation} 
where $\rho = r/a$. It quantifies the overlap between the DE modes and WGMs in the radial direction. It can be estimated by realizing that $\rho^{l_s} \approx \exp( - l_s(1 - \rho))$ for $l_s\gg1$. Therefore, the magnetization of the DE magnon decays rapidly in a reduced length scale of $1/l_s$ (or in a length scale of $a/l_s$). In such a small length, we can approximate WGMs by their value at the surface ($\rho = 1$) giving 
\begin{equation}
	 \mathcal{R} \approx \frac{j_l(k_{\mathbf{p}}a)}{l_s}\left[ j_{l^{\prime}}(k_{\mathbf{q}}a) - j_{l^{\prime}}^{\prime}(k_{\mathbf{q}}a) \right] . 
\end{equation} 
We can use the asymptotic form of the Bessel's function, Eq. (\ref{jlAiry}), along with 
\begin{equation}
	 k_{\mathbf{p}}a = l + \left( \beta_{\nu} - \frac{2^{1/3}P_{\sigma}}{l^{1/3}} \right) \left( \frac{2}{l}\right)^{1/3}, 
\end{equation} 
and the Taylor expansion of the Airy's function around its zeroes for large $l$, 
\begin{equation*}
	 \mathrm{Ai}\left( - \beta_{\nu} + \frac{2^{1/3}P_{\sigma}}{l^{1/3}}\right) \approx \mathrm{Ai}^{\prime}( - \beta_{\nu})\frac{2^{1/3}P_{\sigma}}{l^{1/3}} . 
\end{equation*} 
We can find a similar function for $j_{l^{\prime}}(k_{\mathbf{q}}a)$. We simplify 
\begin{equation}
	 \mathcal{R} \approx \frac{\pi}{4}\left( \frac{4}{ll^{\prime}}\right)^{7/6} \mathrm{Ai}^{\prime}( - \beta_{\nu})\mathrm{Ai}^{\prime}( - \beta_{\nu^{\prime}})P_{TE}(1 + P_{TM}) . 
\end{equation} 
Putting all the constants in Eq. (\ref{Eq:GpmInt}), we arrive at the result mentioned in Eq. (\ref{Coup:DE}). 

\bibliography{References} 

\end{document}